\documentclass[12pt,authoryear]{elsarticle}

\usepackage[hang,flushmargin]{footmisc}

\usepackage{atbegshi,picture}
\usepackage{lipsum}
\usepackage{multirow}
\usepackage{hyperref}

\usepackage{array}
\newcolumntype{C}[1]{>{\centering\arraybackslash}p{#1}}
\newcolumntype{L}[1]{>{\raggedright\arraybackslash}m{#1}}

\usepackage{amssymb}
\usepackage{xcolor}

\usepackage{booktabs}
\graphicspath{{FIGURES/}{TexFiles/}}
\usepackage{subcaption}
\usepackage[section]{placeins}

\usepackage{amssymb}
\usepackage{amsbsy}
\usepackage{amsmath}
\usepackage{booktabs}
\usepackage[figuresleft]{rotating}
\usepackage{longtable}
\usepackage{xcolor}
\usepackage{subcaption}
\usepackage[labelformat=parens,labelsep=quad,skip=3pt]{caption}
\usepackage{graphicx}
\usepackage{soul}
\usepackage{placeins}

\newcommand{\bfx}{\mathbf{x}}
\newcommand{\bfmu}{\boldsymbol{\mu} }
\newcommand{\bfbeta}{\boldsymbol{\beta} }

\newcommand{\bfepsilon}{\boldsymbol{\epsilon} }
\newcommand{\bfgamma}{\boldsymbol{\gamma} }

\newcommand{\bfSigma}{\boldsymbol{\Sigma} }

\newcommand{\bfPhi}{\boldsymbol{\Phi} }

\newcommand{\bfe}{\boldsymbol{e}}

\newcommand{\bfalpha}{\boldsymbol{\alpha}}

\newcommand{\bfz}{\boldsymbol{z}}

\DeclareMathOperator{\diag}{diag}

\usepackage{longtable}

\journal{\footnotesize{Journal of Financial Stability \hspace{0.5cm}}}
\begin{document}

\begin{frontmatter}
\renewcommand{\thefootnote}{\fnsymbol{footnote}}

\title{Systemic Risk in the European Insurance Sector\tnoteref{t1}}
\tnotetext[t1]{We thank Andrea Battista, Riccardo Cesari, Dominick Damast, Stefano Dell'Atti, and participants at the XXVI Workshop on Quantitative Finance (QFW 2025) at the Università degli Studi di Palermo and the 19th International Joint Conference CFE-CMStatistics 2025 for helpful comments. We also acknowledge the comments of the Editor and two anonymous referees. Federico Guerra and Chiara Alvisi provided excellent research assistance. Corresponding author: Nicola Borri (nborri@luiss.it
). All errors are our own.}

\author[Bonaccolto]{Giovanni Bonaccolto}
\author[LUISS1]{Nicola Borri}
\author[Consiglio]{Andrea Consiglio}
\author[LUISS2]{Giorgio Di Giorgio}

\address[Bonaccolto]{Department of Economics and Law, Kore University of Enna, Viale delle Olimpiadi, 94100 Enna, Italy; giovanni.bonaccolto@unikore.it}
\address[LUISS]{Department of Economics and Finance, LUISS University, Viale Romania 32, 00197 Rome, Italy; nborri@luiss.it}
\address[Consiglio]{Department of Economics, Management and Statistics, University of Palermo, Viale delle Scienze, 90128 Palermo, Italy; andrea.consiglio@unipa.it}
\address[LUISS]{Department of Economics and Finance, LUISS University, Viale Romania 32, 00197 Rome, Italy; gdg@luiss.it}

\begin{abstract} 
This paper studies systemic-risk connectedness in the European insurance sector at three levels of granularity: across major segments of financial markets, across insurance subsectors, and across individual insurance companies. Using a common connectedness framework applied to returns, volatility, value-at-risk, and expected shortfall, we document that insurers are an important component of systemic-risk connectedness, especially during stress episodes. We also provide reduced-form evidence on economically relevant channels in the European institutional setting: aggregate insurer spillovers co-move with term spreads, sovereign spreads, and funding stress, and firm-level insurer-to-bank spillovers vary with sovereign risk and domestic sovereign-bond home bias in a way consistent with a balance-sheet channel. The analysis further reveals substantial heterogeneity across subsectors and identifies a stable core of systemically central insurers in firm-level networks.
\end{abstract}



\begin{keyword}
insurance \sep systemic risk \sep GEFVD \sep CAViaR \sep CARES



\end{keyword}

\end{frontmatter}

\section{Introduction}\label{sec:intro}

The insurance market is a crucial component of the financial system. Beyond their traditional role in risk transfer and diversification, insurance companies are major institutional investors, holding large portfolios of government and corporate bonds, equities, and alternative assets. In the aggregate, insurance companies in the European Union held approximately 10 trillion euro in assets at the end of 2024. These portfolios are the ultimate backing for insurers' liabilities, which amounted to approximately 8.5 trillion euro in the same period. In terms of market capitalization, the European insurance sector accounts for about 5\% of broad equity market capitalization, while the banking sector accounts for a further 10\%. Given their size and interconnectedness, insurers are not only exposed to financial market fluctuations and systemic risk, but can also amplify or mitigate financial instability.\footnote{Data on assets and liabilities of European insurance companies are from the European Insurance and Occupational Pensions Authority (EIOPA), available at \href{https://www.eiopa.europa.eu}{www.eiopa.europa.eu}. Data on market capitalization are from Bloomberg.}

This paper quantifies the interconnections between the insurance market and other financial markets from a systemic-risk perspective. We first study relationships between the insurance sector and other key segments of the financial markets---equities, bonds, and banks. We then analyze the insurance sector itself, distinguishing five subsectors---insurance brokers, life and health insurance, multiline insurance, property and casualty insurance, and reinsurance---and their interdependencies. Finally, we examine the network of individual insurance companies.

Our empirical analysis is based on the framework of \cite{Diebold2009,Diebold2014}, applied to a broad set of performance and risk indicators. Rather than proposing a new systemic-risk measure, we use established connectedness tools to compare results across returns, volatility, value-at-risk (VaR), and expected shortfall (ES), and to assess whether conclusions are robust across these dimensions.

The focus on European insurers is motivated by a distinctive institutional setting. Insurers operate under a broadly common regulatory architecture (most notably Solvency II), but face heterogeneous sovereign environments and heterogeneous exposures to domestic government bonds. This differs from much of the U.S.-focused evidence, where insurers are often characterized by common exposure to a single sovereign benchmark and greater emphasis on common corporate and structured-credit exposures \citep{ellul2022insurers,foley2023us}. In Europe, the combination of a common regulatory framework and cross-country variation in sovereign risk provides a useful laboratory to study mechanism-consistent relationships between sovereign stress, portfolio composition, and spillovers.

This setting is economically relevant because insurance business models are naturally exposed to interest-rate and sovereign-risk channels. Insurers typically hold long-duration fixed-income assets to back long-duration liabilities, and many life insurance products embed guarantees that make insurers particularly sensitive to shifts in discount rates and credit conditions. In addition, in several European countries insurance products are distributed through banks, creating links between insurers and the banking system. These features suggest that insurers can act as shock absorbers in some episodes, but also become transmitters of risk when market conditions deteriorate.

Extreme---but not uncommon---events such as the 2007--2009 subprime crisis, the European sovereign debt crisis, and the COVID-19 pandemic illustrate how vulnerabilities in one segment of the financial system can quickly propagate across sectors. During these episodes, insurance firms faced declining asset values, liquidity pressures, and heightened uncertainty, with implications for solvency and market stress. A recent example is the Eurovita crisis in Italy, which followed the rapid increase in interest rates in 2022 and 2023. Although Eurovita was a mid-sized institution, its distress had significant repercussions for the Italian insurance market and for the banks distributing its life insurance products. The episode was contained only after coordinated action by large insurers and distributing banks, supported by public authorities. The Eurovita case highlights how even a relatively small insurer can generate broader financial stress through interconnected portfolios, distribution channels, and confidence effects, underscoring the importance of assessing the systemic role of non-bank financial intermediaries such as insurers.

We collect a comprehensive dataset to study systemic risk and connectedness in the European insurance market. Our sample spans January 3, 2000 to October 22, 2024. First, we use the Euro Stoxx Insurance Equity Index as a proxy for the European insurance sector. This capitalization-weighted index includes insurance stocks from countries in the European Monetary Union (EMU). We also use the Euro Stoxx Banks Index as a proxy for the Euro area banking sector, the EMU Benchmark Government Bond 10-year Index as a proxy for the Euro area government bond market, and the Euro Stoxx Europe 600 ex Financials Index as a proxy for the broader European equity market. Excluding financial firms from the equity benchmark avoids overlap with the insurance and banking indexes. These four indexes serve as our aggregate proxies for the insurance, banking, bond, and equity markets.

Second, we collect data for all publicly listed insurance companies in the Euro area and in geographically proximate regions, including the UK, Switzerland, and Turkey. We use the Bloomberg sectoral classification to assign insurers to five subsectors: insurance brokers, life and health insurance, multiline, property and casualty, and reinsurance. For each subsector, we compute a capitalization-weighted daily log-return series and then estimate time-varying conditional volatility, VaR, and ES.

At the aggregate level, we find a strong comovement between the total spillover index and the insurance sector's ``contribution to others'' across all four indicators. All spillover indices increase markedly during stress episodes. Around the introduction of Solvency II, insurers' contribution declines for several indicators, although this evidence is correlational and may reflect concurrent macro-financial developments. During the COVID-19 pandemic, the insurance sector's contribution to systemic risk rises sharply across all dimensions. After the pandemic, as the ECB began tightening monetary policy in response to high inflation, return-based spillovers decline more clearly, while risk-based spillovers remain more persistent. To provide additional economic interpretation for this time variation, we relate weekly ES-based spillovers from aggregate insurance to macro-financial explanatory variables in the Online Appendix. We find that the term spread, sovereign spread, and funding stress are the most informative variables, while a broad global risk-appetite proxy (the change in the VIX) is not statistically significant.

To further investigate a mechanism suggested by the European institutional setting, we run a firm-level panel exercise on individual insurance companies. Using EIOPA exposures (available starting in 2017Q4), we construct a country-level home-bias measure defined as the total holdings of domestic government bonds by insurers in a given country divided by the total assets of insurers in that country, and interact it with a country-level sovereign spread. In specifications with firm and time fixed effects, and controlling for leverage, the relationship between sovereign stress and insurers' spillovers to banks varies systematically with home bias: the coefficient on sovereign stress is negative, while the interaction between sovereign stress and home bias is positive and statistically significant. This implies that sovereign stress is associated with lower insurer-to-bank spillovers for insurance companies in countries with low home bias, but that this effect weakens and could reverse for insurers in countries with high home bias. We interpret this result as suggestive evidence consistent with a balance-sheet channel, rather than causal identification.

At the insurance subsector level, we uncover substantial heterogeneity. Insurance brokers generally exhibit low ``contribution to others,'' meaning that they are relatively sensitive to shocks from other subsectors but transmit less risk themselves. In contrast, multiline insurers emerge as primary channels through which shocks are propagated, consistent with their broader and more diversified business models. Life and health insurers and property and casualty insurers display intermediate levels of interconnectedness, with ``to'' and ``from'' spillovers that suggest mutual reinforcement. Reinsurers, despite their risk-absorption role, become especially important during stress episodes, when their contribution to aggregate spillovers rises significantly.

We then zoom in to the individual-firm level. By mapping the interconnections among listed insurance companies, we construct weighted and directed networks for each risk indicator. These networks reveal a core cluster of systemically important firms that repeatedly emerges across the return-, volatility-, VaR-, and ES-based networks. Companies such as Aegon, Allianz, Generali, Zurich, Aviva, and Swiss Re are consistently identified as highly central in terms of ``contribution to others,'' suggesting that shocks affecting these firms may be more likely to propagate through the system. The network topology also reveals both subsector and geographic clustering. For example, Turkish insurers tend to cluster together, suggesting the importance of local market or regulatory conditions, while firms from smaller markets (e.g., Slovenian insurers) sometimes form relatively isolated clusters.

We extract central communities from the four firm-level networks using the Louvain algorithm, a widely used modularity-optimization method for community detection \citep{Traag2015}. Modularity measures how well a network is partitioned into groups with dense internal connections and sparse connections across groups. The Louvain algorithm starts with each node as its own community and iteratively reallocates nodes across neighboring communities whenever this increases modularity. Once no further improvement is possible, the network is collapsed into communities and the procedure is repeated on the reduced graph. We focus on the largest community in each network, and the intersection of these largest communities yields a core group of eight insurance companies. Notably, this intersection closely aligns with the list of Global Systemically Important Insurers (G-SIIs) identified by the Financial Stability Board from 2013 to 2016, before the official suspension of that designation.

These results are informative for supervisors and market participants, with important caveats. The evidence supports the use of spillover metrics as monitoring tools and to prioritize where deeper supervisory analysis may be warranted (for example, central firms or subsectors under stress). This is particularly relevant in Europe, where discussions on financial stability continue to address the role of sovereign exposures, the possible development of a common safe asset, and the prudential treatment of government bonds. At the same time, they are market-price-based and reduced-form measures, so we interpret the results as descriptive evidence and mechanism-consistent correlations, to be complemented by balance-sheet and institutional analyses when drawing policy conclusions.

This paper contributes to the literature on systemic risk. \cite{Diebold2009}, \cite{billio2012econometric}, \cite{10.1257/aer.20120555}, \cite{10.1093/rfs/hhw088}, and \cite{brownlees2017srisk} offer alternative frameworks to investigate and quantify systemic risk and market interconnectedness. Most of these studies focus on U.S. non-insurance financial institutions, with the exception of \cite{billio2012econometric}, which considers U.S. insurance companies and finds that banks play a much more critical role in transmitting shocks, and \cite{chodorow2021asset}, which studies U.S. life insurers and argues that their ability to act as ``asset insulators'' allows them to absorb transitory market fluctuations, albeit at the cost of fragility in crisis periods. In addition, \cite{10.1093/rof/rfu012} examines systemic risk among European financial companies using the SRISK model of \cite{brownlees2017srisk}, finding that systemic risk is larger than for U.S. counterparts. Other contributions include \cite{ellul2022insurers}, which argues that U.S.\ life insurers invest in similar high-risk and illiquid bonds, increasing the risk of systemic fire sales in response to common shocks; \cite{foley2023us}, which argues that life insurers are becoming shadow banks through rising exposure to risky corporates and collateralized loan obligations; \cite{annurev:/content/journals/10.1146/annurev-financial-121415-032840}, which emphasizes the role of insurers' corporate bond purchases in stabilizing markets during distress; \cite{MUHLNICKEL2015187}, which studies mergers and systemic risk in insurance; \cite{BERNAL2014270}, which uses CoVaR to compare banks and insurers in the Eurozone and the U.S.; and \cite{BIERTH2015232}, which shows that insurers' systemic risk contribution is generally modest but rises sharply during crises. \cite{borri2022systemic} and \cite{Bonaccolto2023} also use CoVaR to estimate spillovers in European sovereign debt and banking markets, respectively. \cite{foley2020self} theoretically studies self-fulfilling runs in the insurance sector. Relative to this literature, our paper provides a multi-layer analysis of connectedness in the European insurance sector---across markets, subsectors, and firms---using a common framework applied to returns and tail-risk measures, and complements this evidence with a Europe-specific mechanism exercise linking sovereign stress, domestic sovereign-bond home bias, and insurer spillovers.

The rest of the paper is organized as follows. Section \ref{sec:fevd} discusses the network model used to estimate interconnections among financial intermediaries and describes the data. Section \ref{sec:empresults} presents the empirical results. Section \ref{sec:conclusions} concludes.

\section{Methodology and data}\label{sec:fevd}

This section first describes the model used to estimate spillovers in the insurance market. Next, it presents the data used in the empirical analysis.


We investigate spillovers by estimating the Generalized Forecast Error Variance Decomposition (GFEVD) model of \cite{Diebold2009, Diebold2012, Diebold2014}. We extend the GFEVD model to four indicators of risk and performance, and apply it to different degrees of granularity of the insurance market, and to interconnected markets, such as the equity, banking and bond markets. The four indicators we consider are all based on equity prices and are the stock returns and their log-volatility, the value-at-risk (VaR), and the expected shortfall (ES). 
Specifically, volatility is estimated via the Generalized AutoRegressive Conditional Heteroskedasticity (GARCH) model \citep{Bollerslev1986}, VaR via the Conditional AutoRegressive VaR (CAViaR) model \citep{Engle2004}, and ES via the Conditional AutoRegressive ES (CARES) model \citep{Taylor2007}. These four indicators offer a comprehensive framework to analyze the performance and risk in the insurance market. The methodological details regarding the construction of these indicators are provided in Section \ref{sec:metindicators} of the Online Appendix. It is important to clarify the interpretation of the connectedness measures. The GFEVD-based spillover and network statistics used in this paper are reduced-form measures of statistical interdependence, not structural causal effects. They capture the strength and direction of forecast-error linkages across variables, but they do not identify the underlying shocks or the structural mechanisms through which shocks propagate. Accordingly, we interpret the empirical results as evidence on plausible transmission patterns and mechanism-consistent associations, rather than causal effects.

\subsection{Generalized Forecast Error Variance Decomposition}

Let $X_1,\ldots,X_n$ be $n$ real-valued variables, whose realizations at time $t$ are the entries of the $n \times 1$ vector $\bfx_t=\left[x_{1,t} \cdots x_{n,t}\right]^\prime$,  with $t=1,\ldots,T$. 
The Forecast Error Variance Decomposition (FEVD) builds on the following Vector AutoRegressive (VAR) model:
\begin{equation}\label{eq:VAR1}
\bfx_t=\bfalpha+\sum_{i=1}^{p}\bfbeta_i \bfx_{t-i}+\bfepsilon_t,
\end{equation}
where 
$\bfbeta_i$ is an $n \times n$ matrix of parameters that determine the VAR dynamics, $\bfalpha=\left[\alpha_1\cdots \alpha_n\right]^\prime$ is an $n \times 1$ vector of intercepts, and $\bfepsilon_t \sim \mathcal{N} (\textbf{0}, \bfSigma)$ is an $n \times 1$ vector of errors that follow a multivariate Gaussian distribution, such that $\mathbb{E}(\bfepsilon_t \bfepsilon^\prime_s)=0$ and $s \neq t$ \citep{Luetkepohl2007, Pesaran1998}.

The VAR model defined in Equation \eqref{eq:VAR1} has the following infinite moving average representation:
\begin{equation}\label{eq:IMA}
\bfx_t=\bfmu+\sum_{i=0}^{\infty}\bfPhi_i \bfepsilon_{t-i},
\end{equation}
where $\bfmu$ is the unconditional expected value of $\bfx_t$, and $\bfPhi_i$ is obtained from the following recursions: 
\begin{equation}\label{eq:recIMA}
\bfPhi_i = \bfbeta_1 \bfPhi_{i-1} + \bfbeta_2 \bfPhi_{i-2}+\cdots + \bfbeta_p \bfPhi_{i-p},
\end{equation}
with $\bfPhi_0$ being an $n\times n$ identity matrix, and $\bfPhi_i=\textbf{0}$ if $i<0$ \citep{Pesaran1998}.\footnote{In the empirical analysis, we determine the lag order $p$ in Equation \eqref{eq:VAR1} by the Bayesian Information Criterion (BIC).}  

In this paper, we use the Generalized FEVD (GFEVD), proposed by \cite{Pesaran1998}, that, in contrast to the FEVD resulting from orthogonalized impulse response functions, does not depend on the ordering of the variables in $\bfx_t$. In particular, given a forecast horizon $h$, we define the $(i,j)$-th entry of the $n \times n$ GFEVD matrix  as follows:    
\begin{equation}\label{eq:genFEVD}
\theta_{i,j}(h) = \frac{\sigma_{jj}^{-1}\sum_{l=0}^{h} \left(\bfe_i^\prime \bfPhi_l \bfSigma \bfe_j \right)^2}{\sum_{l=0}^{h} \left(\bfe_i^\prime \bfPhi_l \bfSigma \bfPhi_l^\prime \bfe_i \right)},
\end{equation} 
where $\left[\sigma_{11}\cdots \sigma_{nn}\right]=\diag(\bfSigma)$, and $\bfe_j$ is an $n \times 1$ selection vector whose entry $j$ is equal to one, whereas its other elements are zeros \citep{Diebold2014}. 

$\theta_{i,j}(h)$ defined in Equation \eqref{eq:genFEVD} is a measure of spillover from element $j$ to element $i$. More precisely, it is the proportion of the $h$-step ahead forecast error variance of variable $i$ which is accounted for by the innovations in variable $j$, with $i,j=1,\ldots , n$. Note that
$\sum_{j=1}^n \theta_{i,j}(h)$ is not necessarily equal to one for each row $i$ \citep{LN2016, Gross2019, Bax2024}. We then normalize $\theta_{i,j}(h)$ as follows:  
\begin{equation}\label{eq:normgenFEVD}
\tilde{\theta}_{i,j}\left(h\right) = \frac{\theta_{i,j}(h)  }{  \sum_{j=1}^n \theta_{i,j}(h)}\cdot 100,
\end{equation}
for $i=1,\ldots,n$. The coefficients $\tilde{\theta}_{i,j}$ are the central object of estimation and capture the spillovers in the system of equations in the VAR.

Following \cite{Diebold2014}, \cite{LN2016}, \cite{Gross2019}, and \cite{Bax2024}, among others, we compute three different indicators from the $\tilde{\theta}_{i,j}\left(h\right)$ values. First, we refer to the contribution ``from others'' to node $i$: 
\begin{equation}\label{eq:ingoingC}
\tilde{\theta}_{i \leftarrow \bullet}(h) =  \sum_{\substack{j=1 \\ j\neq i}}^{n} \tilde{\theta}_{i,j}\left(h\right).
\end{equation}
Second, we refer to the contribution ``to others'' of node $j$:
\begin{equation}\label{eq:outgoingC}
\tilde{\theta}_{\bullet \leftarrow j}(h) = \sum_{\substack{i=1 \\ i\neq j}}^{n} \tilde{\theta}_{i,j}\left(h\right).
\end{equation}
Finally, we denote the ``total'' spillover index: 
\begin{equation}\label{eq:spillindx}
\tilde{\theta}(h)=\sum_{i=1}^n \sum_{\substack{j=1 \\ j\neq i}}^{n} \tilde{\theta}_{i,j}\left(h\right),
\end{equation}
which captures the sum of all the spillovers across nodes, and thus is a measure of the interconnectedness of a market, or set of nodes, which we also refer to as a ``system''. As with other reduced-form connectedness measures based on market data, these statistics summarize forecast-error variance linkages but do not separately identify direct propagation channels from simultaneous responses to common shocks (e.g., broad interest-rate or risk-appetite factors).

We use the GFEVD model described above to study the presence and propagation of spillovers within three different systems. The first system includes the insurance, equity, banking and bond markets. The second system, which consists of five insurance subsectors (i.e.,  insurance brokers, life health, multiline, property and casualty, and reinsurance), allows the investigation of spillovers across the main subsectors of the insurance market. Because of the relatively small number of variables in $\bfx_t$ (4 and 5, respectively) in both the first and second system, we estimate the parameters of the VAR model in Equation \eqref{eq:VAR1} using the Ordinary Least Squares (OLS) method.

In contrast, the number of estimates $n$ significantly increases in the estimation of the third system, which is the system with the highest degree of granularity in our analysis. In this case, we investigate the network of individual insurance companies, and we address the curse of dimensionality using regularization methods applied to the GFEVD framework. These methods offer the benefit of balancing a slight increase in bias, with a significant reduction in variance \citep{Bonaccolto2023}. In the FEVD and GFEVD framework, \cite{Demirer2017} advocated the Least Absolute Shrinkage and Selection Operator (LASSO) introduced by \cite{Tibshirani1996}, and \cite{Gross2019}, the Elastic Net (ELNET) introduced by \cite{Zou2005}. We choose to apply LASSO, as it provides sparser solutions,  allowing us to identify the most relevant links within the estimated network. However, as discussed by \cite{Fan2001}, LASSO typically provides biased estimates, overshrinking the selected variables.  We follow \cite{Bax2024},  and address this issue by employing the post-LASSO approach described below. 

In a first step of the post-LASSO method,  we LASSO-select the relevant regressors of the $j$-th equation of the VAR model given in \eqref{eq:VAR1} by minimizing the following loss function: 
\begin{equation}\label{eq:lasso}
\mathcal{L}_j  = \sum_{t=p+1}^{T} \left( x_{j,t} - \alpha_j - \sum_{i=1}^{p} \bfbeta_{i[j,:]}  \bfx_{t-i}\right)^2 + \lambda_j  \sum_{i=1}^p \lVert \bfbeta_{i[j,:]}   \rVert_1 ,
\end{equation}
where $\bfbeta_{i[j,:]}$ is the $j$-th row of $\bfbeta_i$,  
$\lVert \bfbeta_{i[j,:]}   \rVert_1$ is the $\ell_1$-norm of $\bfbeta_{i[j,:]}$, and $\lambda_j \geq 0$ is a tuning parameter that determines the sparsity of the solutions: the larger $\lambda_j$, the greater the number of coefficients set to zero \citep{Hastie2009, Murphy2012}, with $j=1,\ldots,n$.\footnote{In our empirical analysis, the value of $\lambda_j$ is determined using five-fold cross-validation.}

We select those regressors whose slope coefficients resulting from the minimization of the loss function in Equation \eqref{eq:lasso} are different from zero. In a second step of the post-LASSO method, we estimate the slope coefficients of the LASSO-selected covariates using the standard OLS estimator, while  the slope coefficients of the regressors which are not LASSO-selected in the first step are set to zero.  



\subsection{Data}\label{sec:data}

We collect a comprehensive dataset to investigate systemic risk and connectedness in the European insurance market. First, we use the Euro Stoxx Insurance Equity Index to proxy for the European insurance market. This is a capitalization-weighted index which includes insurance stocks from countries in the European Monetary Union (EMU). As of March 2025, the Euro Stoxx Insurance equity index has a total market capitalization of approximately 471 billion euro, about 5.6\% of the market capitalization of the aggregate Euro Stoxx Equity Index. The median company in the index has a capitalization of approximately 12 billion euro, and Germany, France and Italy have the highest country weights, respectively equal to 56.5\%, 18.6\% and 10.2\%. Moreover, we use the Euro Stoxx Banks Index, with a market capitalization of approximately 914 billion euro as of March 2025, as representative of the European banking market, and the EMU Benchmark Government Bond 10-year Index as representative of the European government bond market. Furthermore, we use the Stoxx Europe 600 ex Financials Equity Index, which removes financial firms from the Stoxx Europe 600 Index, to capture the residual part of the Euro equity market. This exclusion prevents a potential overlap with the indexes for the insurance and banking markets. We collect daily price data for these four indices from January 3, 2000 to October 22, 2024 and compute log-returns, which are then also used in the estimation of the conditional volatility, value-at-risk and expected shortfall. 

Furthermore, we collect daily stock prices for all publicly listed insurance companies in the European Economic Area, as well as in geographically proximate countries such as the United Kingdom, Switzerland, and Turkey.\footnote{Specifically, we select our sample using the equity screening function in Bloomberg and applying filters to include all listed insurance companies in Eastern and Western Europe. We acknowledge that focusing only on listed companies provides a broad but not fully representative sample of the insurance sector, especially for countries with significant unlisted insurance companies. Implicitly, we assume that the behavior of listed and unlisted insurers is broadly similar, an assumption that appears reasonable both theoretically (given similar objective functions) and empirically (given broadly comparable business practices).} From these prices, we compute daily log-returns. The resulting dataset is unbalanced, as some insurance companies entered the market after the beginning of our sample period (January 3, 2000), while others exited before its end (October 22, 2024). 

We adopt the Bloomberg's sectoral classification to assign insurance companies to one of the following five subsectors: insurance brokers, life and health insurance, multiline, property and casualty, and reinsurance.\footnote{The Bloomberg classification corresponds to the field \textit{INDUSTRY\_SUBGROUP}, which assigns each stock based on the primary source of its total revenues. For example, a company is classified under the life and health insurance sector if a majority of its revenues originates from life and health insurance activities.} Table \ref{tab:infocompanies} in Section \ref{app:tablesfigures} of the Online Appendix reports the complete list of the insurance companies included in our dataset along with their subsector classification. Figure \ref{fig:activepositions}, also in Section \ref{app:tablesfigures} of the Online Appendix,  plots the evolution in the number of insurance companies clustered by subsector for which daily log-returns are available. The figure indicates a general positive trend in the number of insurance companies. In particular, the number of companies in the insurance brokers subsector increases from 2 to 13, with a median value of 7. For the other subsectors, we observe the following initial and final number of insurance companies: 9 and 32 for life and health insurance; 20 and 47 for multiline; 8 and 37 for property and casualty; 4 and 10 for reinsurance. 

Moreover, for each of the five subsectors of the insurance market, We compute the daily logged return as the weighted average of the log-returns of the insurance companies within that subsector. Specifically, the daily weight of each company within a given subsector is determined by its market capitalization on day $t$, divided by the total market capitalization of all companies in the same subsector on the same day. In case a given company presents missing values on a given day $t$, we set its daily logged return and weight equal to zero. From the logged return time series, we also estimate the time-varying conditional volatility, value-at-risk, and expected shortfall time series using the GARCH, CAViaR, and CARES models. 

Table \ref{tab:summarylogret} reports descriptive statistics about the data used in the empirical analysis. Panel A refers to the main market indices, while Panel B to the insurance stock-level data. The summary statistics for the main market indices indicate differences in return characteristics across asset classes. The mean log-returns (in percentages) vary, with the Insurance Index (0.002) and Bank Index (-0.012) exhibiting lower values compared to Stocks ex-Financials (0.009) and Government Bonds (0.006). The standard deviation is highest for the banking (1.899) and insurance (1.731) equity indices, while government bonds exhibit the lowest volatility (0.370). The 5th and 95th percentiles highlight the extent of downside and upside risks, with the banking market displaying the widest dispersion (-2.927 to 2.777). For the insurance stock-level data in Panel B, the equally-weighted broad (All) mean return is high (0.012), and higher than the mean return on the main market indices. This indicates that small capitalization insurance stocks performed better than large capitalization insurance stocks, with the caveat that the Euro Stoxx Insurance Equity Index include only EMU stocks. For the insurance sub-sectors, the highest return is for insurance brokers (0.047), followed by property and casualty. The insurance brokers subsector has also the highest volatility (1.882) among the five subsectors. Finally, for the insurance stock-level data, we also report the number of stocks in each sub-sector. The sub-sectors with the largest number of components, in our sample, are multiline (38) and property and casualty (26), while insurance brokers (7) and reinsurance (7) are the sub-sectors with the smallest number of components. 

\begin{table}[ht]
\centering
\resizebox{\textwidth}{!}{
\begin{tabular}{llllllll}
  \toprule
 & Mean & Std & Q5 & Q50 & Q95 & N & T \\ 
  \midrule
  \multicolumn{8}{c}{\textbf{Panel A: main market indices}} \\
Insurance Equity Index & 0.002 & 1.731 & -2.609 & 0.023 & 2.487 & 1 & 6472 \\ 
  Bank Equity Index & -0.012 & 1.899 & -2.927 & 0.007 & 2.777 & 1 & 6472 \\ 
  Stocks ex Fin. Equity Index & 0.009 & 1.119 & -1.798 & 0.031 & 1.658 & 1 & 6472 \\ 
  Govt Bond Index & 0.006 & 0.370 & -0.596 & 0.007 & 0.560 & 1 & 6472 \\
  \midrule
  \multicolumn{8}{c}{\textbf{Panel B: insurance stock-level data}} \\
  Insurance Brokers & 0.047 & 1.882 & -2.319 & 0.000 & 2.493 & 7 & 6472 \\ 
  Life Health & 0.010 & 1.528 & -2.380 & 0.039 & 2.160 & 20 & 6472 \\ 
  Multiline & 0.013 & 1.412 & -2.097 & 0.039 & 1.961 & 38 & 6472 \\ 
  Property and Casualty & 0.026 & 1.564 & -2.295 & 0.053 & 2.175 & 26 & 6472 \\ 
  Reinsurance & 0.012 & 1.626 & -2.342 & 0.045 & 2.207 & 7 & 6472 \\ 
  All & 0.012 & 2.375 & -3.253 & 0.004 & 3.295 & 70 & 5584 \\ 
   \bottomrule
\end{tabular}
}
\caption{This table presents summary statistics for the logged return series (in percentages). Panel A refers to the main market indices: the equity indices for insurance, banking, non-financial stocks, and government bonds. Panel B refers to insurance stock-level data. Specifically, the table reports the mean, standard deviation, and the 5th, 50th, and 95th percentiles. For the insurance subsectors we consider value-weighted indices, while ``All'' refers to simple average across all the individual insurance stocks. The number of individual assets (N) and daily observations (T) are also reported. The sample period spans from January 3, 2000, to October 22, 2024. Data source: Datastream.} 
\label{tab:summarylogret}
\end{table}

\begin{figure}[hbt!]
\hspace{-0.4cm}
\includegraphics[scale=0.16]{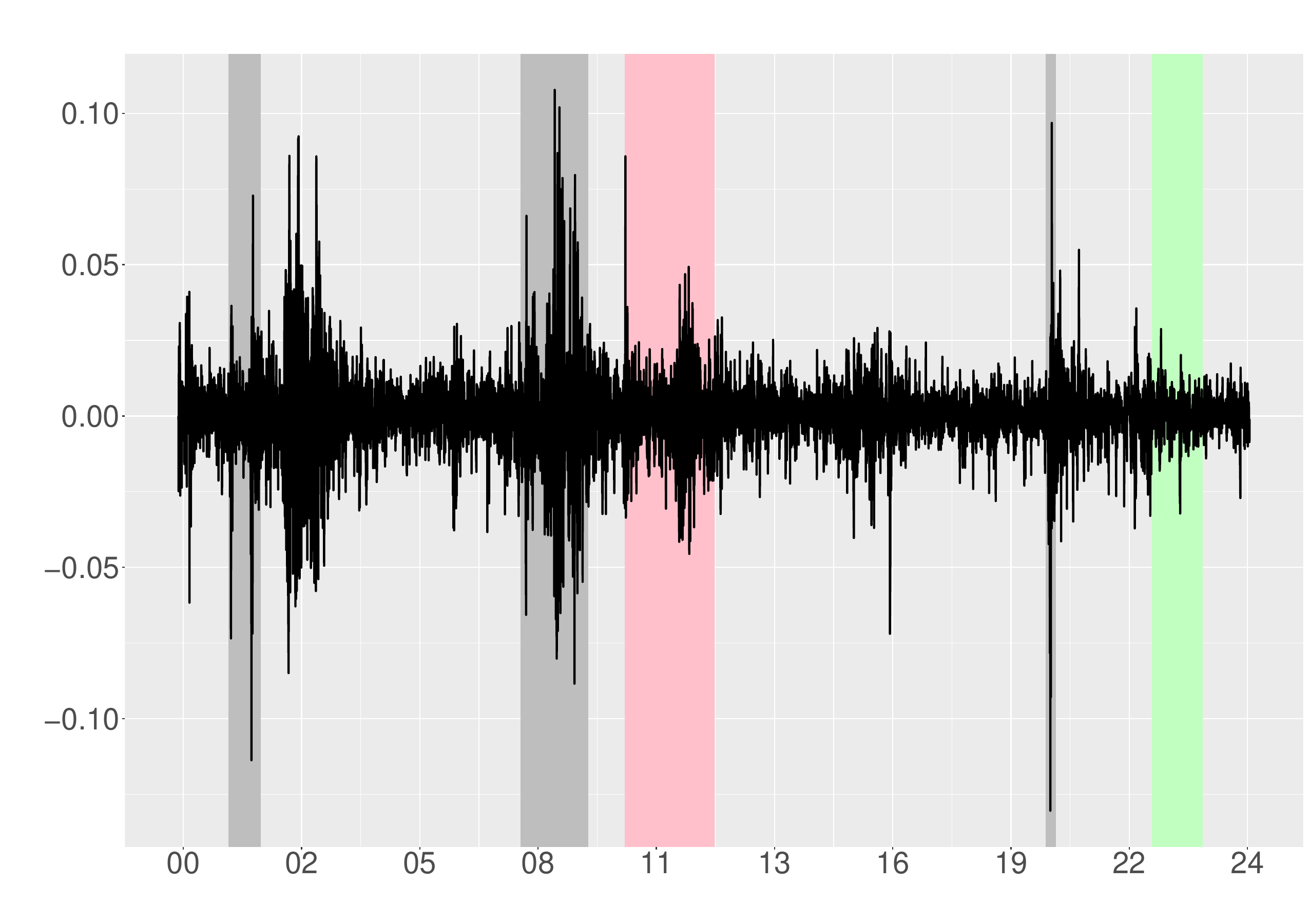}
\includegraphics[scale=0.16]{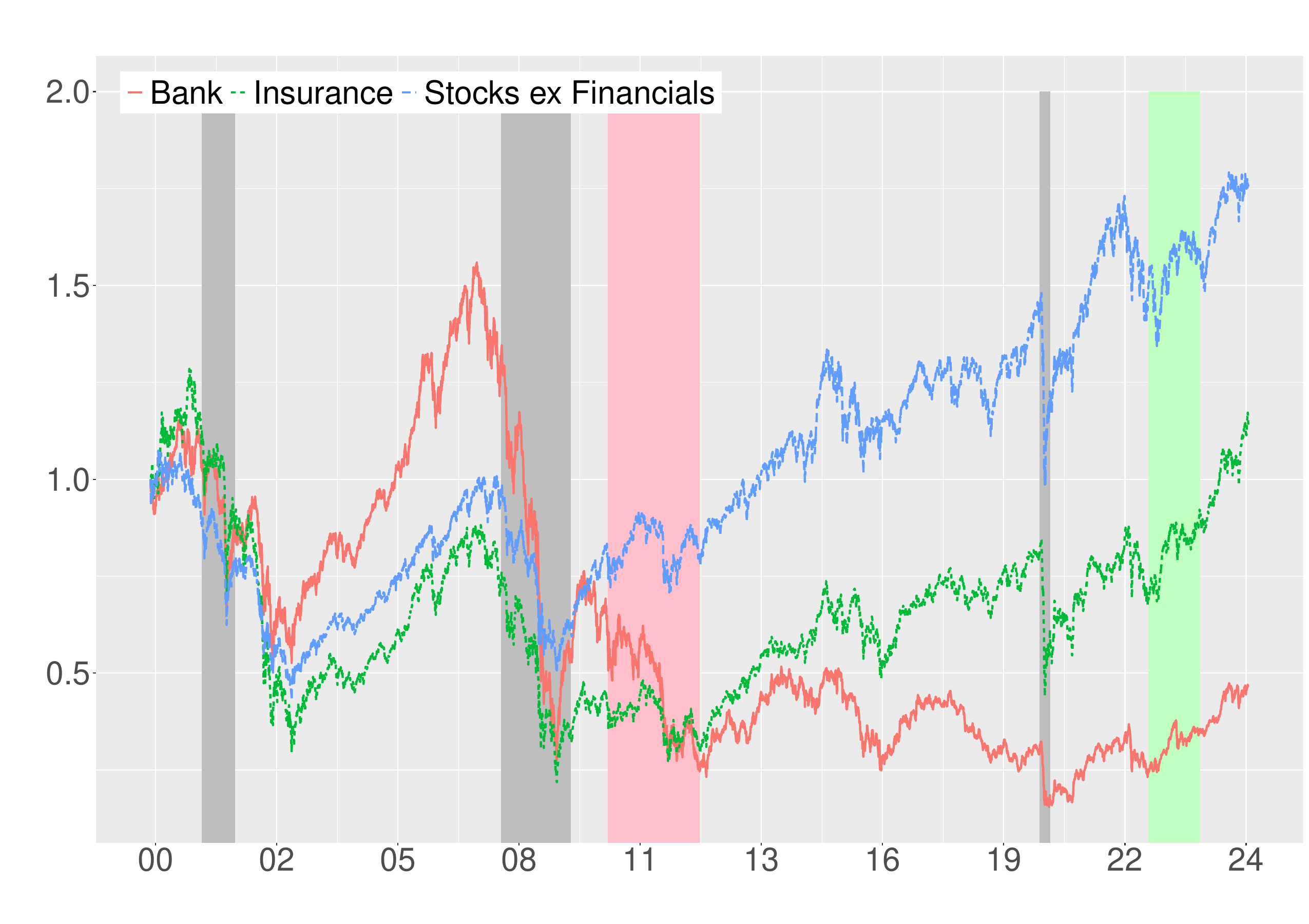}
\caption{This figure plots the logged daily return time series for the Euro Stoxx Insurance equity index (left plot) and the cumulated return indices for the Euro Stoxx Insurance equity index (green line), together with the banking (red line) and equity ex-financial indices (blue line). The gray shaded bars correspond to NBER recessions, the pink shaded bar corresponds to the European sovereign debt crisis, from May 2010 to June 2012, and the green shaded bar refers to the ECB interest rate increase, from July 2022 to September 2023. The sample is from January 3, 2000 to October 22, 2024. Data are from Datastream.}
    \label{fig:insurance_logreturns}
\end{figure}

Figure \ref{fig:insurance_logreturns} plots the daily log-return associated with the Euro Stoxx Insurance Equity Index (left subplot) and the cumulated returns of the main equity indices (right plot). The figure highlights two interesting stylized facts. First, the daily log-returns of the insurance equity index reveal heteroskedasticity and volatility clustering \citep{Cont2001}, and the frequency of large tail events, such as the COVID-19 pandemic, which produced the largest daily loss of $-17.4\%$ on March 12, 2020. This evidence motivates the construction, from the log-returns, of time-varying additional measures of risk, such as the conditional volatility, value-at-risk, and expected shortfall. Second, the cumulated return series of the three main equity indices reveals the high co-movement, but also marked difference in performance over the sample considered. Moreover, the figure reveals the particularly high correlation, across these markets, of the largest tail events.

\section{Empirical analysis}\label{sec:empresults}

This section presents the empirical results from the estimation of the GFEVD model. We first present the results for the insurance, banking, equity, and government bond markets in Section \ref{sec:macrosectors}. Section \ref{sec:sovrisk} then presents a firm-level panel analysis of sovereign risk, home bias, and insurer spillovers in the European institutional setting, while Section \ref{sec:solvency} provides a descriptive comparison of spillover patterns around the introduction of Solvency II. Section \ref{sec:subsectors} considers spillovers across insurance subsectors, while Section \ref{sec:companies} presents the findings for the network of individual insurance companies. Finally, Section \ref{sec:robustness} presents additional results and extensions.

We estimate the model both unconditionally, using the full-sample, as well as conditionally, with dynamic estimates based on rolling windows. For the dynamic estimates, we consider a rolling window size of 250 daily observations (i.e., approximately one year) and a one-day forward step. 

Following \cite{Gross2019} and \cite{Bax2024}, we set the forecast horizon $h$ in Equation \eqref{eq:genFEVD} equal to 10, which is a reasonable choice from a risk management viewpoint \citep{Diebold2014}. We determine the lag order $p$ on the basis of the Bayesian Information Criterion (BIC) computed on the underlying VAR model. The selected $p$ values vary according to the system of interest and the performance or risk measure that is adopted to estimate the GFEVD model. Specifically, in the system comprising the insurance, banking, bond, and equity markets, the chosen values of $p$ are 1 (GFEVD on log-return time series), 3 (conditional log-volatility), 1 (CAViaR), and 5 (CARES). In the case of the five insurance subsectors, the corresponding $p$ values are 1, 1, 1, and 3, respectively. Finally, for individual insurance companies, $p$ is always set to 1 across all estimations.

\subsection{Insurance and other financial markets}\label{sec:macrosectors}

In this section, we present the results obtained from the estimation of the GFEVD model at the most aggregated level, investigating spillovers across the insurance, banking, government bond, and equity markets. Specifically, we consider spillovers using the log-return, the conditional log-volatility, the CAViaR, and the CARES.

\begin{figure}[hbt!]
\hspace{0cm}
\includegraphics[scale=0.33]{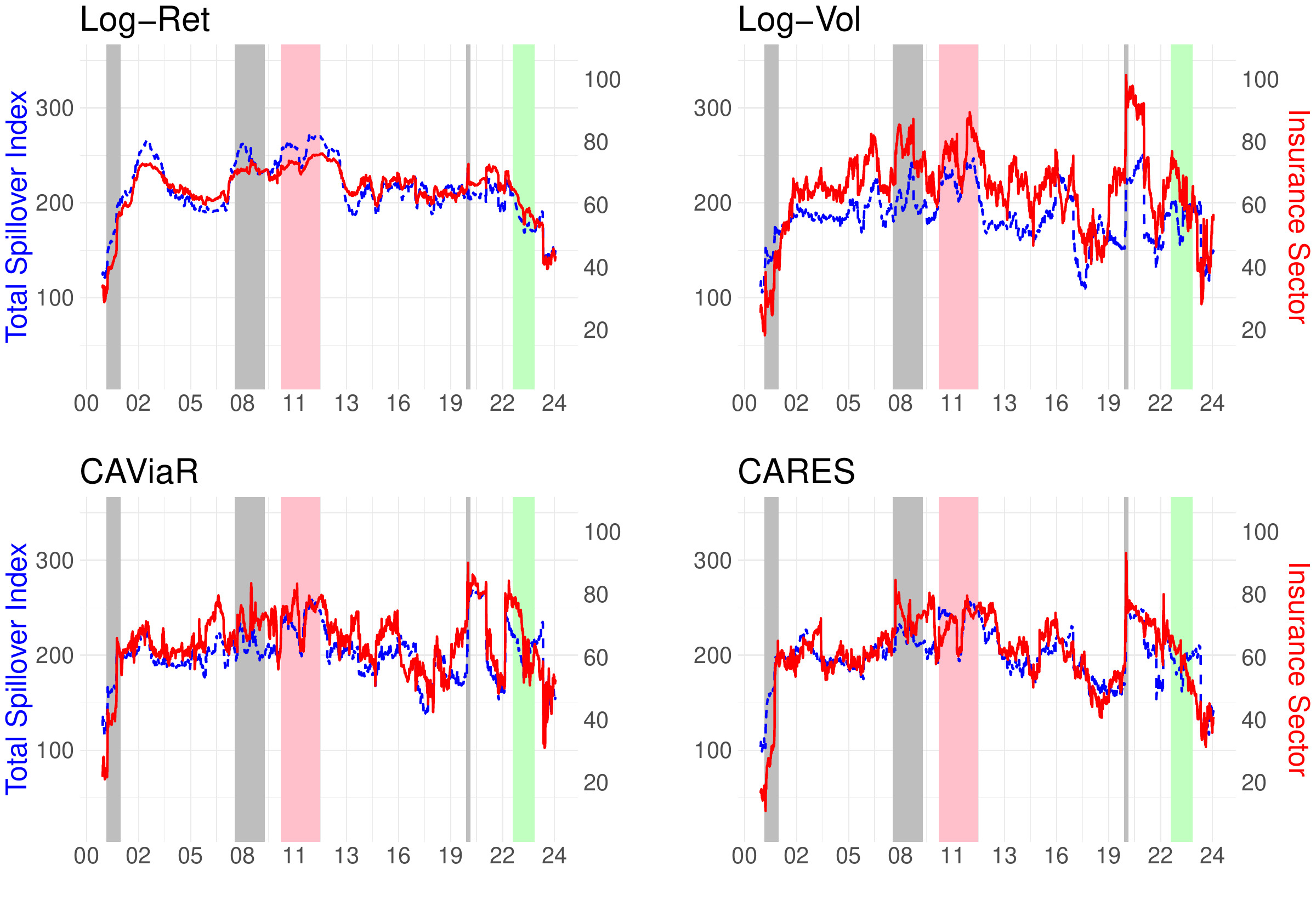}
\caption{This figure plots the evolution of the contribution to others of the European insurance sector (red-solid line), as defined in Equation \eqref{eq:outgoingC}, and of the total spillover index (blue-dashed line), as defined in Equation \eqref{eq:spillindx}, resulting from the estimation of the GFEVD model on the log-return (top-left panel), conditional log-volatility (top-right panel), CAViaR (bottom-left panel), and CARES (bottom-right panel) time series. 
The total spillover index is represented on the left vertical axis, while the contribution to others of the insurance sector is represented on the right vertical axis. The dynamic estimation is based on a 250-day rolling window and 1-day ahead forecast. The gray shaded bars correspond to NBER recessions, the pink shaded bar corresponds to the European sovereign debt crisis, from May 2010 to June 2012, and the green shaded bar refers to the ECB interest rate increase, from July 2022 to September 2023. The sample goes from January 3, 2000 to October 22, 2024.}
\label{fig:TotSpill_Insur}
\end{figure}

Figure \ref{fig:TotSpill_Insur} summarizes the results of the dynamic spillover estimates. It shows the evolution of the total spillover index (blue-dashed line, left vertical axis) across the insurance, banking, bond, and equity markets, for each of the four performance and risk indicators. In parallel, the figure plots the evolution of the insurance sector's ``contribution to others'' index (red-solid line, right vertical axis), capturing the extent to which shocks originating in the insurance market spill over to the remaining markets. To support the interpretation, the figure includes vertical colored bands that highlight key economic episodes: gray bands correspond to NBER U.S. recession periods (specifically, the dot-com bubble, the subprime crisis, and the COVID-19 pandemic); the pink band marks the European sovereign debt crisis (May 2010 to June 2012); and the green band denotes the period of ECB monetary tightening (July 2022 to September 2023).

A first observation is the strong comovement between the total spillover index and the insurance sector's contribution to others across all four indicators. The unconditional correlation coefficients are 0.90, 0.75, 0.81, and 0.92, respectively, pointing to a tight link between overall systemic risk and the insurance sector's role in its propagation. Notably, all spillover measures tend to rise sharply during the major economic stress episodes highlighted in the figure, such as the NBER recessions. We emphasize that this comovement may reflect both interconnectedness/propagation and common exposure to dominant shocks, so we interpret it as reduced-form evidence rather than a structural decomposition of transmission channels.

Second, the figure shows that all spillover indices increased markedly following the end of the 2001 recession (associated with the dot-com bubble) and have remained at structurally higher levels since, with only a modest decline toward the end of the sample. This suggests a long-term shift in the systemic interconnectedness of financial markets, which could be due to the relevant changes in technology, internet development and digitalization that made much more rapid the responses of agents and markets to new information.

A closer comparison between the total spillover index and the insurance sector spillover index reveals further insights. In relative terms (i.e., relative to their own vertical scales in Figure \ref{fig:TotSpill_Insur}), spillovers in returns remained lower for the insurance sector compared to the total spillover index until the European sovereign debt crisis. Since then, however, the insurance sector's spillovers have generally remained relatively more elevated. In contrast, for all risk indicators, the insurance-sector index appears more elevated relative to its own scale throughout most of the sample, highlighting the sector's comparatively strong role in risk transmission. Because the two series are plotted on different vertical axes, these comparisons should be interpreted as relative comovement patterns rather than absolute magnitudes; in absolute terms, the total spillover index is always larger.

Toward the end of the sample, an interesting divergence emerges across the different spillover indicators. During the COVID-19 pandemic, the insurance sector's contribution to systemic risk spiked sharply across all dimensions. While spillovers based on returns gradually declined after the crisis, those derived from risk measures---namely volatility, value-at-risk, and expected shortfall---exhibited a more complex and persistent pattern. Specifically, all three risk-based indicators show that both the total spillover index and the insurance sector index rose substantially during the pandemic, with the latter increasing relatively more. This elevated spillover from the insurance sector is consistent with significant stress in the industry and stronger co-movement with other markets during the episode, although the reduced-form connectedness measures cannot separately identify direct transmission from common shocks.

Following the immediate pandemic shock, spillover indices for risk measures temporarily receded but surged again when the ECB began raising interest rates in response to unexpectedly high inflation. These rate hikes disproportionately affected the insurance sector, given its exposure to long-duration fixed-income assets, whose valuations declined as interest rates rose. As inflationary pressures later eased and market expectations shifted toward lower future rates, the spillover indices gradually declined, approaching levels observed at the start of the sample period.

These findings underscore the importance of looking beyond return-based measures when analyzing systemic risk. While return spillovers suggest a steady post-pandemic decline, risk-based indicators reveal a more nuanced and dynamic picture of financial contagion, highlighting the value of incorporating higher-order risk metrics in systemic risk assessments.

\begin{table}[htbp]
\begin{footnotesize}
  \centering
    \begin{tabular}{lC{1.5cm}C{1.6cm}C{1.6cm}C{1cm}C{1.4cm}C{1.4cm}}
    \toprule
    Period & Start & End & Days  & Total & Insurance \\
    \midrule
    \multicolumn{6}{c}{Panel A: Log-Return} \\
    \midrule
    Dot-com bubble & 01/03/2001 & 30/11/2001 & 197   & 171.36 & 44.83 \\
    Sub-prime crisis & 03/12/2007 & 30/06/2009 & 412   & 245.35 & 71.17 \\
    EU Sovereign debt crisis & 03/05/2010 & 01/06/2012 & 545   & 258.88 & 73.29 \\
    COVID-19 & 03/02/2020 & 30/04/2020 & 64    & 209.69 & 66.83 \\
    ECB interest rate increase & 21/07/2022 & 21/09/2023 & 306   & 182.00 & 59.16 \\
    Normal times & 03/01/2000       & 22/10/2024      & 4699  & 211.55 & 64.76 \\
    \midrule
    \multicolumn{6}{c}{Panel B: Log-Volatility} \\
    \midrule
    Dot-com bubble & 01/03/2001 & 30/11/2001 & 197   & 151.07 & 33.45 \\
    Sub-prime crisis & 03/12/2007 & 30/06/2009 & 412   & 203.05 & 76.89 \\
    EU Sovereign debt crisis & 03/05/2010 & 01/06/2012 & 545   & 226.91 & 76.46 \\
    COVID-19 & 03/02/2020 & 30/04/2020 & 64    & 195.52 & 85.08 \\
    ECB interest rate increase & 21/07/2022 & 21/09/2023 & 306   & 187.45 & 65.00 \\
    Normal times & 03/01/2000      & 22/10/2024      & 4699  & 179.78 & 63.69 \\ 
    \midrule
    \multicolumn{6}{c}{Panel C: CAViaR} \\    
    \midrule
    Dot-com bubble & 01/03/2001 & 30/11/2001 & 197   & 174.29 & 45.78 \\
    Sub-prime crisis & 03/12/2007 & 30/06/2009 & 412   & 211.43 & 70.87 \\
    EU Sovereign debt crisis & 03/05/2010 & 01/06/2012 & 545   & 237.05 & 74.17 \\
    COVID-19 & 03/02/2020 & 30/04/2020 & 64    & 230.14 & 78.15 \\
    ECB interest rate increase & 21/07/2022 & 21/09/2023 & 306   & 206.95 & 66.16 \\
    Normal times &03/01/2000       &  22/10/2024     & 4699  & 197.09 & 63.64 \\  
    \midrule
    \multicolumn{6}{c}{Panel D: CARES} \\   
    \midrule
    Dot-com bubble & 01/03/2001 & 30/11/2001 & 197   & 163.92 & 35.70 \\
    Sub-prime crisis & 03/12/2007 & 30/06/2009 & 412   & 218.67 & 72.27 \\
    EU Sovereign debt crisis & 03/05/2010 & 01/06/2012 & 545   & 238.54 & 70.47 \\
    COVID-19 & 03/02/2020 & 30/04/2020 & 64    & 220.55 & 71.98 \\
    ECB interest rate increase & 21/07/2022 & 21/09/2023 & 306   & 190.95 & 58.89 \\
    Normal times &   03/01/2000    &  22/10/2024     & 4699  & 195.56 & 60.73 \\    
    \bottomrule
    \end{tabular}%
  \caption{This table reports summary statistics of systemic risk spillovers across sub-samples and different performance and risk indicators. Panel A to D refer to log-return, log-volatility, CAViaR and CARES, respectively. For each sub-sample, the table reports the start and end dates, the number of trading days within each period, the mean value of the total spillover index, and the mean contribution to others of the insurance sector. The dynamic GFEVD is estimated using a 250-day rolling window with a step of one day ahead.}
  \label{tab:total&insurance_specialperiods}
  \end{footnotesize}
\end{table}

The evidence discussed above is summarized in Table \ref{tab:total&insurance_specialperiods}, where we report the average values of the two indicators shown in Figure \ref{fig:TotSpill_Insur} across different periods of financial and economic stress. The table presents results for the different performance and risk indicators, namely log-returns (Panel A), log-volatility (Panel B), CAViaR (Panel C), and CARES (Panel D).

Several patterns emerge from the data. First, the total spillover index exhibits significant increases during crisis periods relative to normal times across all four indicators. This pattern is consistent with heightened interconnectedness of financial markets during periods of systemic distress. Second, the insurance sector's contribution to spillovers also rises substantially in these periods, particularly for risk-based indicators (Panels B to D). Notably, during the COVID-19 crisis, the insurance sector's contribution to total spillovers reached its highest levels across all risk measures, underscoring the sector's heightened vulnerability and systemic importance in times of severe financial turmoil.

The ECB's interest rate hikes present a different dynamic. While the total spillover index increased compared to normal times, its rise was more contained than in previous crises, likely reflecting a more gradual market adjustment. However, the insurance sector's contribution remained elevated, particularly for log-volatility and CAViaR, highlighting the sector's sensitivity to interest rate changes due to its exposure to long-duration fixed-income assets. As financial conditions stabilized following the peak of inflationary pressures, spillover indices began to decline, nearing levels observed in normal times. Overall, these findings emphasize the need to analyze spillovers across multiple dimensions of risk beyond returns, as systemic stress can manifest more prominently through volatility, tail risk, and systemic fragility measures.

\begin{figure}[hbt!]
\hspace{-0.4cm}
\includegraphics[scale=0.34]{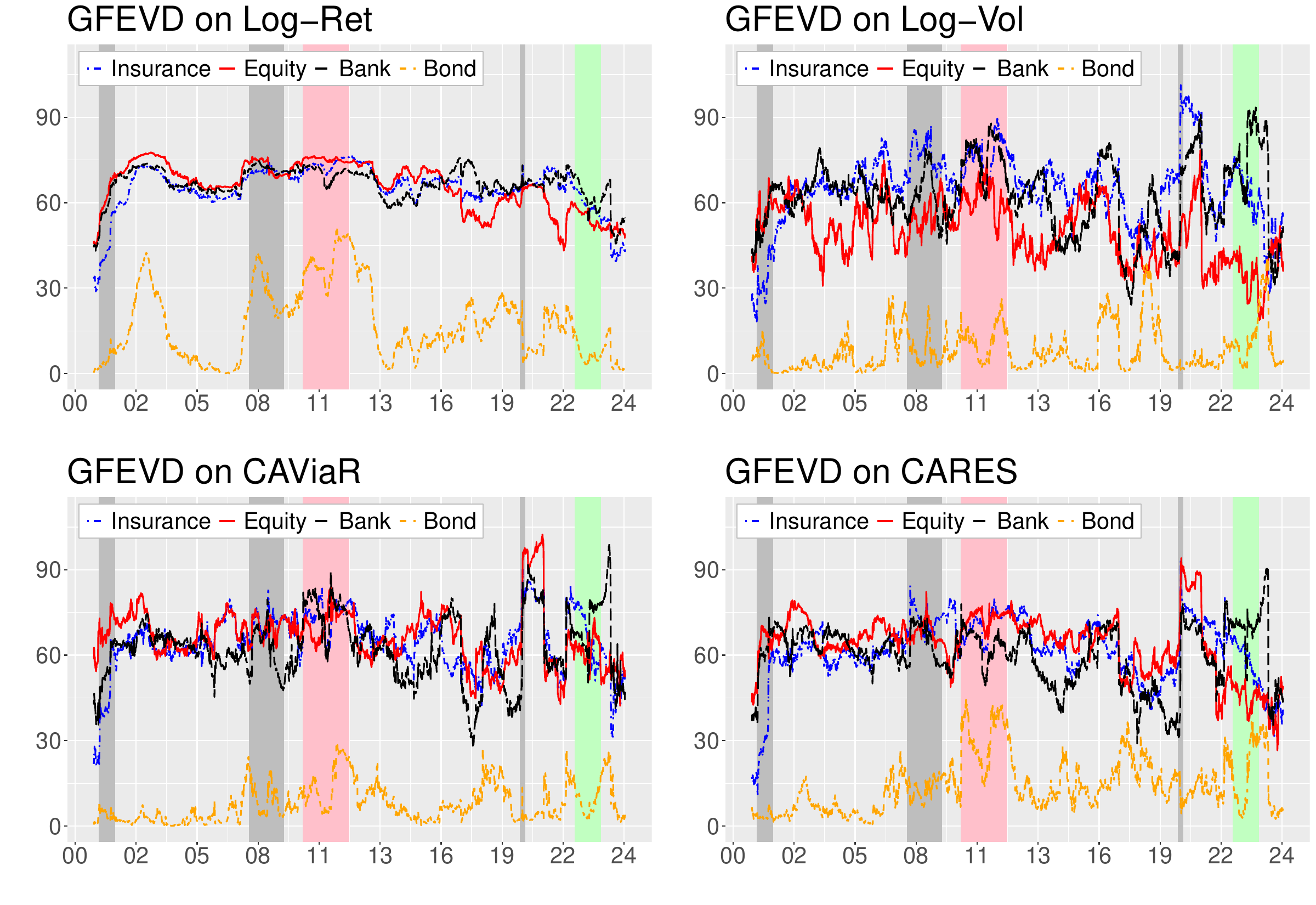}
\caption{This figure plots the evolution of the spillover index for the broad equity, insurance, banking, and bond markets. The four panels correspond to estimation of the GFEVD model on log-returns (top left), conditional volatility (top right), CAViaR (bottom left) and CARES (bottom right). The dynamic estimation is based on a 250-day rolling window and 1-day ahead forecast. The gray shaded bars correspond to NBER recessions, the pink shaded bar corresponds to the European sovereign debt crisis, from May 2010 to June 2012, and the green shaded bar refers to the ECB interest rate increase, from July 2022 to September 2023. The sample goes from January 3, 2000 to October 22, 2024.}
\label{fig:MacroSectors_SystemicImpact}
\end{figure}

So far, we have compared the dynamics of the total spillover index and the impact of the insurance sector. We now directly contrast the systemic importance of the insurance sector with that of other financial markets—banking, bond, and equity markets—as illustrated in Figure \ref{fig:MacroSectors_SystemicImpact}. Once again, we observe pronounced spikes in systemic contributions during major financial crises, with these spikes becoming more pronounced as we shift from log-returns (top-left panel) to risk-based measures.

A key takeaway from the figure is that spillover measures based on log-returns exhibit more muted fluctuations compared to those based on risk indicators, underscoring the importance of looking beyond return-based spillovers to assess systemic risk. Across all periods, the insurance sector consistently maintains a high contribution to systemic risk, comparable to that of the banking and equity markets. The bond market, on the other hand, has a relatively lower systemic impact during stable periods but sees a sharp increase during financial crises, such as the US subprime crisis.

In the most recent period, characterized by the ECB's interest rate hikes, we observe a divergence in systemic contributions across sectors. While the insurance, banking and government bond markets experienced a substantial increase in their contributions, coinciding with tighter financial conditions, the contributions of the equity markets ex-financials declined. This divergence highlights sector-specific sensitivities to rising interest rates: insurance companies, banks and bonds are more directly exposed to changes in monetary policy and yield curves, while the rest of the equity market---despite being typically described as a long-duration asset---saw a relative decline in systemic impact.

\subsection{Sovereign Risk, Home Bias, and Insurer Spillovers}\label{sec:sovrisk}

The previous sections document that insurers are systemically relevant in Europe and that their spillovers are strongly time-varying. We now move from description to mechanism. The key economic question is whether sovereign-risk shocks transmit differently across insurers depending on portfolio composition. This is a particularly relevant test in Europe, where sovereign risk is heterogeneous across countries and insurers differ substantially in their exposure to domestic government bonds \citep{IMF2018EAInsurance}. This institutional setting contrasts with the U.S., where insurers hold claims on a common sovereign benchmark (U.S. Treasuries), which mechanically limits cross-country variation in sovereign-credit exposure; consistent with this contrast, U.S.-focused studies emphasize common exposures to corporate-credit and structured-credit markets rather than cross-country sovereign-credit heterogeneity \citep{ellul2022insurers,foley2023us}.

\begin{figure}[htbp]
\centering
\includegraphics[width=\textwidth]{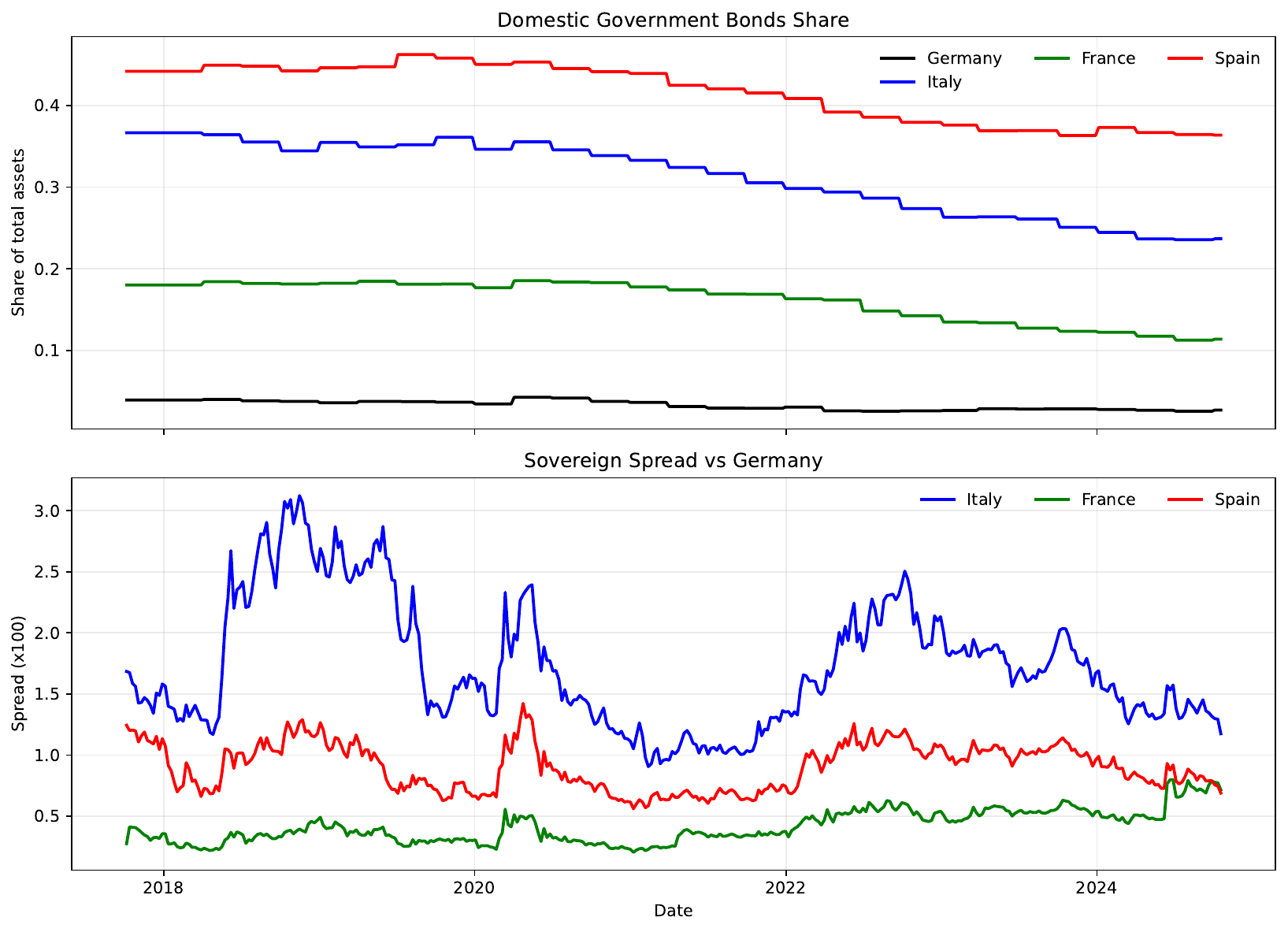}
\caption{This figure plots the evolution of domestic government bond share and sovereign spread for insurance companies in Germany, Italy, France, and Spain. Top panel: share of domestic government bonds in insurers' total assets (\(HomeBias\)). Bottom panel: sovereign spread relative to Germany (\(SovSpread\)). Data are country-level and from EIOPA exposure statistics and Bloomberg. Sovereign spreads are for the 10-year maturity. The sample starts in 2017Q4.}
\label{fig:homebias_sovspread_countries}
\end{figure}

To measure this channel, we use country-level exposures from EIOPA. Specifically, for each country and quarter, starting in 2017Q4, we compute \(HomeBias\) as total holdings of domestic government bonds by insurers in that country divided by total insurer assets in that country. This is the same home-bias variable used in the panel regressions below. Figure \ref{fig:homebias_sovspread_countries} shows the evolution of this measure together with sovereign spreads for four representative countries (Germany, Italy, France, and Spain). At the beginning of the sample, insurers' domestic government-bond share ranges from below 5\% in Germany to above 40\% in Spain. Although these shares decline over time, sizable cross-country heterogeneity remains at the end of the sample, with Spain still above 35\% and Italy around 25\%. Over the same period, Italy displays the highest sovereign spread in the group, declining from about 3\% to around 1\%.

The empirical design exploits two features of the European setting. First, sovereign stress is not uniform across countries, so the spread over Germany varies meaningfully over time and across insurers' home jurisdictions. Second, insurers exhibit heterogeneous home-bias intensity in government-bond portfolios. If this balance-sheet channel is relevant, the effect of sovereign stress on insurer spillovers should depend on \(HomeBias\), captured by the interaction \(HomeBias \times SovSpread\).

Table \ref{tab:twopanel_logret_cares} reports pooled estimates and estimates with company and time fixed effects for spillovers from insurers to banks, bonds, and equities. All panel specifications include a measure of leverage \(Lev\), from Bloomberg, to control for company-specific time-varying riskiness. The pooled specification combines within-company and cross-company variation. By contrast, company fixed effects absorb time-invariant insurer characteristics (e.g., business model and persistent risk profile), while time fixed effects absorb shocks common to all firms (e.g., broad macro-financial conditions). Relative to pooled estimates, the fixed-effects specification isolates within-insurer variation over time, net of aggregate shocks.

Our interpretation focuses on the bank-spillover equations in the specification with company and time fixed effects. In Panel A (CARES), the coefficient on \(SovSpread\) is negative and significant (\(-2.543\)), while the interaction \(HomeBias \times SovSpread\) is positive and significant (\(3.936\)). In Panel B (Log-Ret), the same pattern holds: \(SovSpread=-1.154\) (significant) and \(HomeBias \times SovSpread=2.321\) (significant). Thus, higher sovereign stress is associated with lower insurer-to-bank spillovers for insurance companies in countries with low home bias, but this effect weakens and can reverse for insurers in countries with higher home bias. The implied sign-switch thresholds, $-\beta_{SovSpread}/\beta_{HomeBias\times SovSpread}$, are approximately $0.646$ in Panel A (CARES) and $0.497$ in Panel B (Log-Ret). In the estimation sample used for Table~\ref{tab:twopanel_logret_cares}, the home-bias measure ranges from $0.000$ to $0.583$ (with $p_{95}=0.522$ and $p_{99}=0.583$). Thus, the threshold lies outside the observed support in Panel A (CARES), while in Panel B (Log-Ret) it lies within the upper tail of the sample.

Outside banks, coefficients are less stable in the fixed-effects specification, consistent with a bank-centered sovereign-risk transmission channel. This evidence complements the aggregate time-series results reported in the Online Appendix (Table \ref{tab:ins_spillovers_econ_conditions}), where we relate weekly ES-based spillovers from aggregate insurance to macro-financial explanatory variables. In those regressions, the term spread, sovereign spread, and funding stress are the most informative variables, while a broad global risk-appetite proxy (the change in the VIX) is not statistically significant. As a robustness check, Appendix Table~\ref{tab:twopanel_logret_cares_SE_country_clustered} reports the same panel regressions with standard errors clustered at the country level, which is a relevant alternative because the key explanatory variables (sovereign spreads and home bias) vary at the country level; the CARES results remain qualitatively consistent, while the Log-Ret estimates are less precisely estimated (though with similar signs), as expected given the limited number of country clusters in the sample.

\begin{table}[!htbp]
\centering
\small
\resizebox{\textwidth}{!}{%
\begin{tabular}{lcccccc}
\hline
 & \multicolumn{3}{c}{Pooled} & \multicolumn{3}{c}{TWFE} \\
 & Bank & Bond & Equity & Bank & Bond & Equity \\
\hline
\multicolumn{7}{l}{\textbf{Panel A: CARES}} \\
Lev & 0.342*** & 0.081*** & 0.307*** & 0.015 & -0.039 & -0.031 \\
 & (0.122) & (0.031) & (0.104) & (0.086) & (0.025) & (0.074) \\
Home Bias & 22.075*** & 5.381*** & 16.889*** & -10.667 & -9.880 & -16.157 \\
 & (4.578) & (1.089) & (4.094) & (32.131) & (7.149) & (27.659) \\
Sov Spread & 2.059** & 0.812*** & 3.577*** & -2.543*** & 0.136 & -0.566 \\
 & (0.803) & (0.213) & (0.875) & (0.814) & (0.189) & (0.906) \\
Home Bias $\times$ Sov Spread & -5.629*** & -1.877*** & -8.136*** & 3.936** & -0.240 & 0.926 \\
 & (1.757) & (0.451) & (1.940) & (1.550) & (0.396) & (1.670) \\
N & 33882 & 33882 & 33882 & 33882 & 33882 & 33882 \\
Firm FE & No & No & No & Yes & Yes & Yes \\
Time FE & No & No & No & Yes & Yes & Yes \\
\hline
\multicolumn{7}{l}{\textbf{Panel B: Log-Ret}} \\
Lev & 0.372** & 0.097** & 0.313** & -0.072 & -0.025 & -0.063 \\
 & (0.147) & (0.045) & (0.126) & (0.091) & (0.047) & (0.048) \\
Home Bias & 24.331*** & 6.656*** & 23.822*** & 14.784 & 8.231 & 31.109 \\
 & (6.796) & (1.703) & (5.803) & (24.664) & (6.300) & (19.708) \\
Sov Spread & 2.427*** & 0.851*** & 4.024*** & -1.154* & -0.268 & -0.429 \\
 & (0.849) & (0.305) & (0.945) & (0.685) & (0.318) & (0.698) \\
Home Bias $\times$ Sov Spread & -6.544*** & -1.786*** & -9.963*** & 2.321* & 0.560 & 0.697 \\
 & (1.957) & (0.675) & (2.161) & (1.243) & (0.631) & (1.269) \\
N & 33236 & 33236 & 33236 & 33236 & 33236 & 33236 \\
Firm FE & No & No & No & Yes & Yes & Yes \\
Time FE & No & No & No & Yes & Yes & Yes \\
\hline
\end{tabular}
}%
\caption{This table reports panel-regression estimates of directional spillovers from insurers to Bank, Bond, and Equity markets. The dependent variable is the directional spillover from insurance to each destination market. Columns labeled TWFE report two-way fixed-effects estimates (firm and time fixed effects). Robust standard errors are one-way clustered at the firm level and reported in parentheses. *** $p<0.01$, ** $p<0.05$, * $p<0.10$.}

\label{tab:twopanel_logret_cares}
\end{table}

\subsection{Spillover Patterns Around the Introduction of Solvency II}\label{sec:solvency}

During our sample period, a major regulatory reform of the European insurance sector came into effect: Solvency II, introduced on January 1, 2016. Solvency II reshaped the prudential framework for insurers by tightening capital requirements, introducing risk-sensitive solvency ratios, and strengthening risk management and disclosure standards. Because our framework considers different measures of performance and risk, it is well suited to documenting whether insurers' role in risk transmission changed around this regulatory transition. We emphasize, however, that this is a descriptive comparison: changes between 2015 and 2016 may also reflect contemporaneous macro-financial developments, including the ECB's monetary policy stance, changes in bank-resolution expectations, the Brexit referendum, and shifts in sovereign-risk perceptions.

Table~\ref{tab:contribution_to_others_solvency2} reports systemic risk contributions before (2015) and after (2016) the introduction of Solvency II. Between 2015 and 2016, the contribution of insurers to other sectors declined across log-volatility and the tail-risk measures (CAViaR and CARES), while remaining broadly stable in returns. By contrast, contributions from banks and bonds increased substantially, particularly under the tail-sensitive indicators, while equities displayed a more mixed pattern. These patterns are consistent with a relative decline in insurers' measured contribution to tail spillovers in the immediate post-reform period, but they should not be interpreted as causal effects of Solvency II.

More generally, these results illustrate the usefulness of our econometric framework for tracking how spillover patterns evolve around major regulatory and macro-financial events, especially when focusing on volatility and tail-risk measures rather than returns alone. Table \ref{tab:contribution_to_others_multi} in the Online Appendix shows that the same qualitative patterns are broadly similar when we use three-year windows before and after 2016.

\begin{table}[htbp]
  \centering
  \begin{footnotesize}
    \begin{tabular}{L{2cm}C{2.4cm}C{2.4cm}C{2.4cm}C{2.4cm}}    
    \toprule
\multicolumn{5}{c}{Contribution to others} \\
\midrule
 & Log-Return & Log-Volatility & CAViaR & CARES \\
\midrule
\multicolumn{5}{c}{Panel A: Year 2015} \\
\midrule
Insurance & 67.53 & 70.22 & 72.14 & 71.72 \\
Equity    & 62.04 & 55.57 & 66.51 & 53.39 \\
Bank      & 63.28 & 46.51 & 49.99 & 51.71 \\
Bond      &  8.82 &  5.57 &  2.65 &  9.05 \\
\midrule
\multicolumn{5}{c}{Panel B: Year 2016} \\
\midrule
Insurance & 68.92 & 57.93 & 53.65 & 61.33 \\
Equity    & 52.39 & 63.06 & 67.44 & 52.96 \\
Bank      & 68.55 & 73.02 & 72.11 & 70.80 \\
Bond      & 13.44 & 19.35 & 10.34 & 20.96 \\
\midrule
\multicolumn{5}{c}{Panel C: Percentage Variation (\%)} \\
\midrule
Insurance &   2.05 & -17.50 & -25.63 & -14.49 \\
Equity    & -15.56 &  13.48 &   1.38 &  -0.80 \\
Bank      &   8.32 &  57.00 &  44.23 &  36.92 \\
Bond      &  52.28 & 247.17 & 290.19 & 131.62 \\
\bottomrule
\end{tabular}
 \caption{Systemic risk contributions around the introduction of Solvency II. This table reports the contribution to others of the insurance, equity, bank and bond markets estimated in 2015 (Panel A) and in 2016 (Panel B). The contribution to others is reported for the log-return, the log-volatility, the CAViaR and the CARES risk measures. The insurance market regulation Solvency II was introduced January 1, 2016. Panel C reports the percentage variation in the contribution to others between 2015 and 2016.}
\label{tab:contribution_to_others_solvency2}
  \end{footnotesize}
\end{table}%

\subsection{The subsectors of the insurance market}\label{sec:subsectors}

In this section, we present the results from the estimation of the GFEVD model applied to the different insurance subsectors: insurance brokers, life and health, multiline, property and casualty, and reinsurance. For each sector, we construct a value-weighted portfolio containing the insurance stocks using the sector-categorization from Bloomberg. 

We start with the analysis of the results from the unconditional estimates, based on the full-sample, and summarized in Table \ref{tab:spilltable_fs_subsectors}. The four panels of the table refer to the spillovers in terms of returns (Panel A), conditional volatility (Panel B), CAViaR (Panel C) and CARES (Panel D).

\begin{table}[htbp]
  \centering
  \begin{footnotesize}
    \begin{tabular}{C{2.8cm}C{1cm}C{1cm}C{1cm}C{1cm}C{1cm}|C{3cm}}    
    \toprule
          & \multicolumn{6}{c}{From} \\
\cmidrule{2-7}    To    & Ins.Bro. & Lif.Hea. & Mul.Lin. & Pro.Cas. & \multicolumn{1}{c}{Reins.} & \textit{Contr. from others} \\
    \midrule
    \multicolumn{7}{c}{Panel A: Log-Return} \\
    \midrule
    Ins.Bro. & 79.23 & 4.39  & 7.24  & 4.13  & 5.01  & \textit{20.77} \\
    Lif.Hea. & 2.29  & 36.06 & 23.74 & 19.37 & 18.54 & \textit{63.94} \\
    Mul.Lin. & 3.13  & 21.32 & 32.73 & 20.48 & 22.34 & \textit{67.27} \\
    Pro.Cas. & 2.17  & 19.78 & 23.38 & 36.72 & 17.95 & \textit{63.28} \\
    Reins. & 2.47  & 18.60 & 24.89 & 17.52 & 36.52 & \textit{63.48} \\
    \midrule
    \textit{Contr. to others} & \textit{10.07} & \textit{64.09} & \textit{79.24} & \textit{61.50} & \textit{63.84} & \textit{\textbf{Spillover Index:}} \\
    \textit{Net Contribution} & \textit{-10.70} & \textit{0.15} & \textit{11.97} & \textit{-1.78} & \textit{0.36} & \textit{\textbf{278.74}} \\
    \midrule 
    \multicolumn{7}{c}{Panel B: Log-Volatility} \\
    \midrule
    Ins.Bro. & 94.51 & 1.30  & 1.86  & 1.50  & 0.83  & \textit{5.49} \\
    Lif.Hea. & 0.59  & 42.43 & 24.04 & 16.81 & 16.13 & \textit{57.57} \\
    Mul.Lin. & 0.59  & 20.40 & 40.42 & 18.37 & 20.23 & \textit{59.58} \\
    Pro.Cas. & 0.53  & 16.92 & 22.24 & 45.27 & 15.04 & \textit{54.73} \\
    Reins. & 0.36  & 16.30 & 24.21 & 15.20 & 43.93 & \textit{56.07} \\
    \textit{Contr. to others} & \textit{2.07} & \textit{54.91} & \textit{72.35} & \textit{51.88} & \textit{52.22} & \textit{\textbf{Spillover Index:}} \\
    \textit{Net Contribution} & \textit{-3.42} & \textit{-2.66} & \textit{12.77} & \textit{-2.85} & \textit{-3.85} & \textit{\textbf{233.44}} \\ 
    \midrule
    \multicolumn{7}{c}{Panel C: CAViaR} \\
    \midrule
    Ins.Bro. & 87.44 & 2.88  & 4.51  & 2.61  & 2.56  & \textit{12.56} \\
    Lif.Hea. & 0.69  & 35.55 & 26.04 & 18.71 & 19.01 & \textit{64.45} \\
    Mul.Lin. & 0.76  & 20.98 & 35.77 & 20.26 & 22.23 & \textit{64.23} \\
    Pro.Cas. & 0.65  & 17.91 & 24.97 & 38.76 & 17.72 & \textit{61.24} \\
    Reins. & 0.37  & 17.91 & 25.70 & 17.45 & 38.57 & \textit{61.43} \\
    \textit{Contr. to others} & \textit{2.47} & \textit{59.67} & \textit{81.22} & \textit{59.02} & \textit{61.52} & \textit{\textbf{Spillover Index:}} \\
    \textit{Net Contribution} & \textit{-10.09} & \textit{-4.78} & \textit{16.99} & \textit{-2.22} & \textit{0.09} & \textit{\textbf{263.90}} \\ 
    \midrule  
    \multicolumn{7}{c}{Panel D: CARES} \\    
    \midrule
    Ins.Bro. & 84.24 & 3.16  & 4.28  & 5.51  & 2.81  & \textit{15.76} \\
    Lif.Hea. & 1.62  & 38.19 & 24.38 & 17.34 & 18.47 & \textit{61.81} \\
    Mul.Lin. & 1.92  & 20.24 & 36.17 & 20.28 & 21.39 & \textit{63.83} \\
    Pro.Cas. & 2.38  & 15.51 & 23.55 & 41.24 & 17.32 & \textit{58.76} \\
    Reins. & 1.43  & 17.16 & 24.82 & 18.38 & 38.21 & \textit{61.79} \\
    \textit{Contr. to others} & \textit{7.35} & \textit{56.07} & \textit{77.03} & \textit{61.52} & \textit{60.00} & \textit{\textbf{Spillover Index:}} \\
    \textit{Net Contribution} & \textit{-8.41} & \textit{-5.74} & \textit{13.20} & \textit{2.76} & \textit{-1.79} & \textit{\textbf{261.95}} \\    
    \bottomrule
    \end{tabular}%
  \caption{This table summarizes the spillover matrix across the European insurance subsectors estimated by the GFEVD model. Furthermore, for each subsector, it reports the contribution to others, the contribution from others, and the net contribution. Finally, the table reports the total spillover index. Panel A to D refer to log-return, log-volatility, CAViaR and CARES, respectively. The model is estimated for the following subsectors: insurance brokers (Ins.Bro.), life health (Lif.Hea.), multiline (Mul.Lin.), property and casualty (Pro.Cas.), and reinsurance (Reins.). The sample goes from January 3, 2000 to October 22, 2024.}
  \label{tab:spilltable_fs_subsectors}
  \end{footnotesize}
\end{table}%

A joint examination of the four spillover matrices in Table \ref{tab:spilltable_fs_subsectors} shows distinct patterns in the transmission of systemic risk across insurance subsectors. A robust pattern is the prominent role of the multiline insurance subsector, which consistently generates high spillovers---measured by $\tilde{\theta}_{i,j}(h)$---to all other subsectors except insurance brokers. Insurance brokers appear to be the most isolated segment, exhibiting low spillover values both in terms of contributions received and transmitted. Excluding brokers, the spillovers from multiline to other subsectors range from 22.24 (to property and casualty in Panel B) to 26.04 (to life and health in Panel C), significantly exceeding the spillovers originating from other subsectors. The systemic relevance of multiline insurers confirms results for an earlier sample of global financial institutions in \cite{kaserer2019systemic}.

The systemic relevance of the multiline subsector is further highlighted by its contributions to others, $\tilde{\theta}_{\bullet \leftarrow j}(h)$, where it is the only subsector consistently exceeding a value of 70. Notably, it reaches its peak at 81.22 in Panel C (CAViaR). The contrast between multiline and the other subsectors becomes even more pronounced when considering net contributions, calculated as the difference between contributions to and from others, $\tilde{\theta}_{\bullet \leftarrow j}(h) - \tilde{\theta}_{j \leftarrow \bullet}(h)$. While all other subsectors either exhibit negative values or remain close to zero, multiline consistently registers strongly positive net contributions, ranging from 11.97 in Panel A (log-returns) to 16.99 in Panel C (CAViaR). These results underscore the central role of the multiline subsector in the systemic risk dynamics of the insurance market.

These findings suggest that both balance sheet size and the nature of the business model---particularly the distinction between financial and non-financial activities---are relevant for the transmission of systemic risk. In this context, the broad diversification typical of multiline insurers, as well as their role (shared with reinsurers) as ultimate risk absorbers through reinsurance, appears to have a limited mitigating effect. The large balance sheets of these subsectors, rather than diversification per se, appear more closely associated with their central position in our spillover measures.

Turning to the total spillover index, $\tilde{\theta}(h)$, the highest value of 278.74 is observed when estimating the GFEVD model on log-return time series (Panel A), indicating that return-based spillovers are more pronounced compared to risk-based measures.


\begin{figure}[hbt!]
\hspace{-0.2cm}
\includegraphics[scale=0.34]{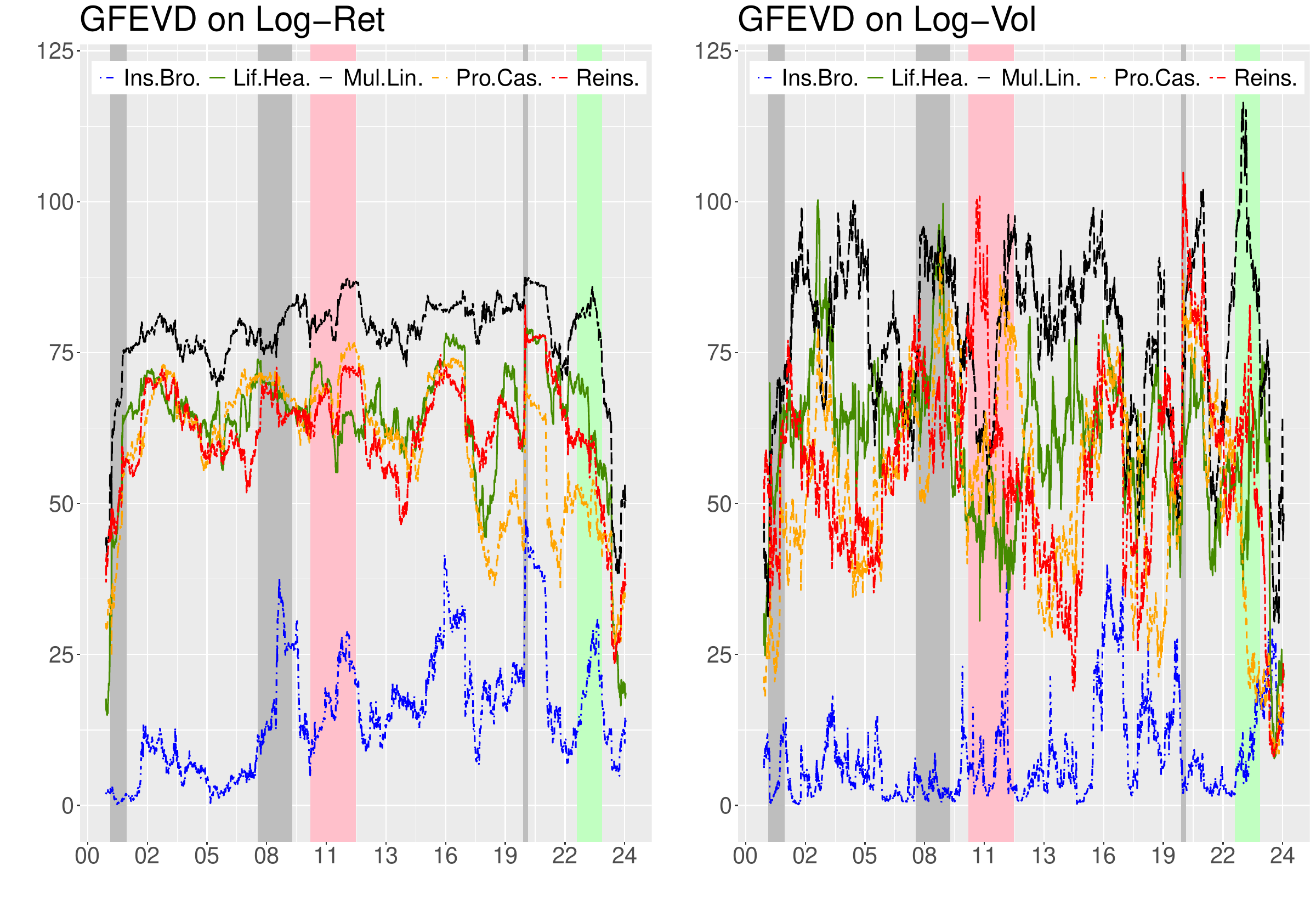}
\caption{This figure plots the evolution of the impact (quantified by the contribution to others) of the insurance brokers (Ins.Bro.), life health (Lif.Hea.), multiline (Mul.Lin.), property and casualty (Pro.Cas.), and reinsurance (Reins.) subsectors, resulting from GFEVD estimated on the log-return (left panel) and conditional log-volatility (right panel) time series. The dynamic estimation is based on a 250-day rolling window and 1-day ahead forecast. The gray shaded bars correspond to NBER recessions, the pink shaded bar corresponds to the European sovereign debt crisis, from May 2010 to June 2012, and the green shaded bar refers to the ECB interest rate increase, from July 2022 to September 2023. The sample goes from January 3, 2000 to October 22, 2024.}
\label{fig:SubSectorsRollingImpacts_RetVol}
\end{figure}

After analyzing the unconditional estimates, we now focus on the dynamic estimates of the spillover for the five subsectors of the insurance market. Figure \ref{fig:SubSectorsRollingImpacts_RetVol} summarizes the results, plotting the dynamic estimates for log-returns (left panel) and conditional log-volatilities (right panel).\footnote{The results obtained from the CAViaR and CARES time series are qualitatively similar and available upon request.} 

Figure \ref{fig:SubSectorsRollingImpacts_RetVol} shows that the impacts of the subsectors are not constant over time, but are affected by the important events that occurred during the analyzed period, and that we cited above. The estimates obtained from log-returns appear more stable and smoothed, whereas those derived from log-volatilities are more volatile and exhibit significant spikes, which often coincides with the relevant events highlighted with the colored bars.

The dynamic estimates are consistent with the earlier finding that the multiline subsector (black line) has the largest spillover impact in the insurance market. In particular, the contribution to others of the multiline subsector is persistently above that of the remaining subsectors, and is largest during the US subprime financial crisis and the COVID-19 pandemic. Interestingly, during the COVID-19 pandemic related economic shock, the relative importance of the reinsurance sector increased sharply and, in the case of the spillovers in terms of log-volatility, reached a high-value above 100. Finally, the insurance brokers' sector contribution to risk is the lowest, although it also rises sharply in periods of financial and economic distress.

\subsection{Insurance companies}\label{sec:companies}

In this section, we consider spillovers in the insurance market using a granular sample containing individual insurance stocks. 

To address the issue of an unbalanced sample of insurance stocks, we implement a two-step strategy. First, we replace missing values with zeros, implicitly assuming that an insurance stock did not contribute to spillovers on those particular days. This assumption is consistent with the idea that if no data is available for a company on a given day, it had no significant market impact or did not trigger any notable events. Second, we filter time series by retaining only those in which nonzero log-returns exceed 30\% of observations, corresponding to at least 1,941 trading days. This criterion ensures that the presence of zero log-returns does not unduly distort the results. A total of 70 companies meet this requirement, as indicated in Table \ref{tab:summarylogret} and by the value ``1'' in the fourth column of Table \ref{tab:infocompanies} in \ref{app:tablesfigures}.  

We first estimate the GFEVD model for the 70 companies we selected. Next, we investigate their interconnections and spillovers using a directed network. Note that, because of the large number of parameters of the model estimated on individual stocks, we estimate the VAR parameters using the post-LASSO regression method.

\begin{figure}[hbt!]
\hspace{-5.5cm}
\includegraphics[scale=0.8]{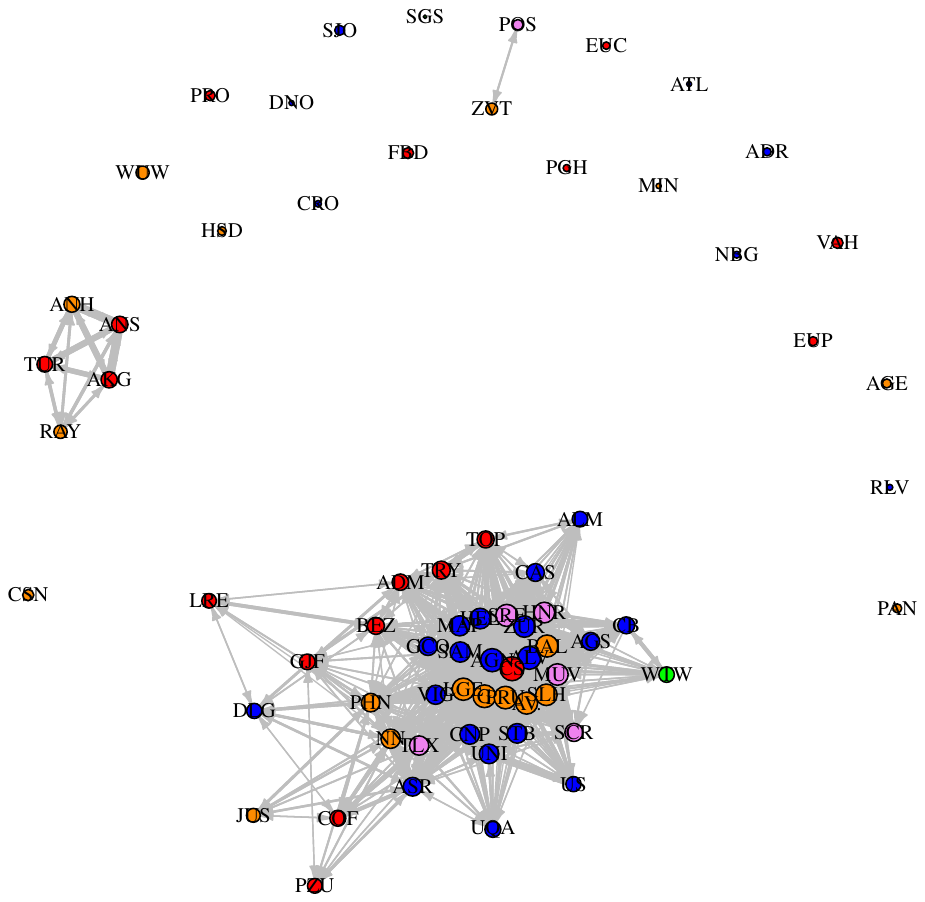}
\vspace{-2cm}
\caption{Network of insurance companies, color-coded by subsector—insurance brokers (green), life and health (orange), multiline (blue), property and casualty (red), and reinsurance (violet)—based on the GFEVD model estimated on the logged return time series.}
\label{fig:CompaniesNetwork_LogRet}
\end{figure}

\begin{figure}[hbt!]
\hspace{-5.5cm}
\includegraphics[scale=0.8]{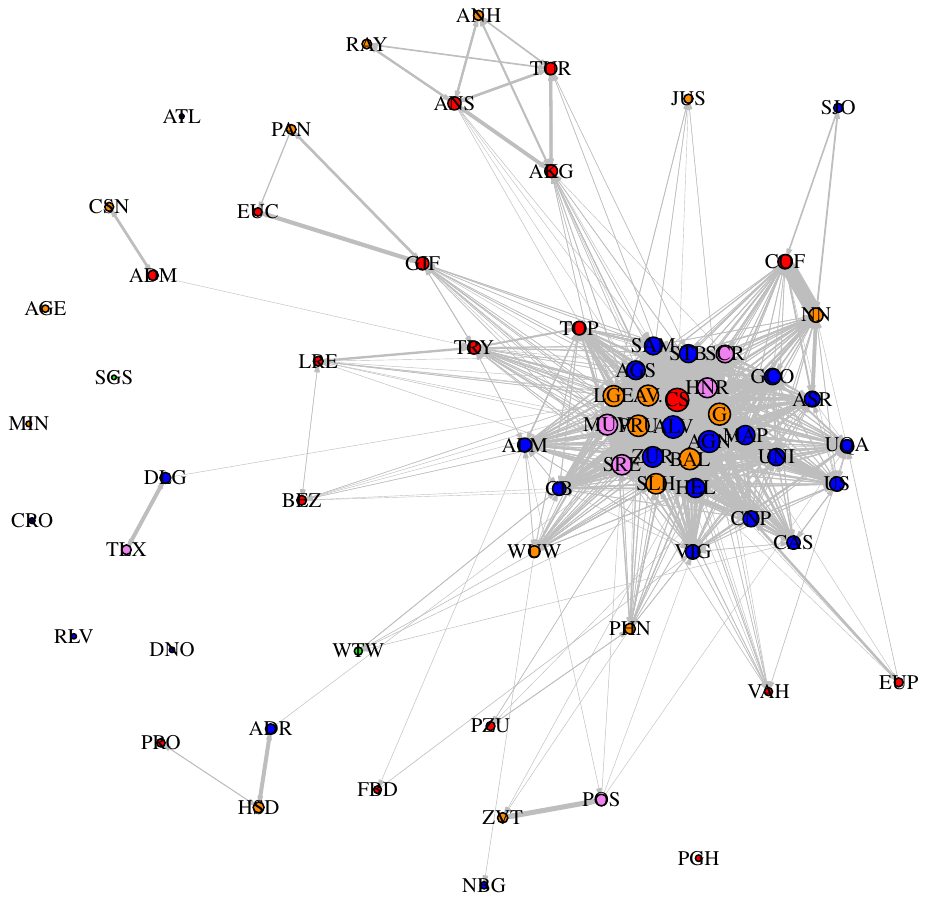}
\vspace{-2cm}
\caption{Network of insurance companies, color-coded by subsector—insurance brokers (green), life and health (orange), multiline (blue), property and casualty (red), and reinsurance (violet)—based on the GFEVD model estimated on the logged conditional volatility time series.}
\label{fig:CompaniesNetwork_LogVol}
\end{figure}

\begin{figure}[hbt!]
\hspace{-5.5cm}
\includegraphics[scale=0.8]{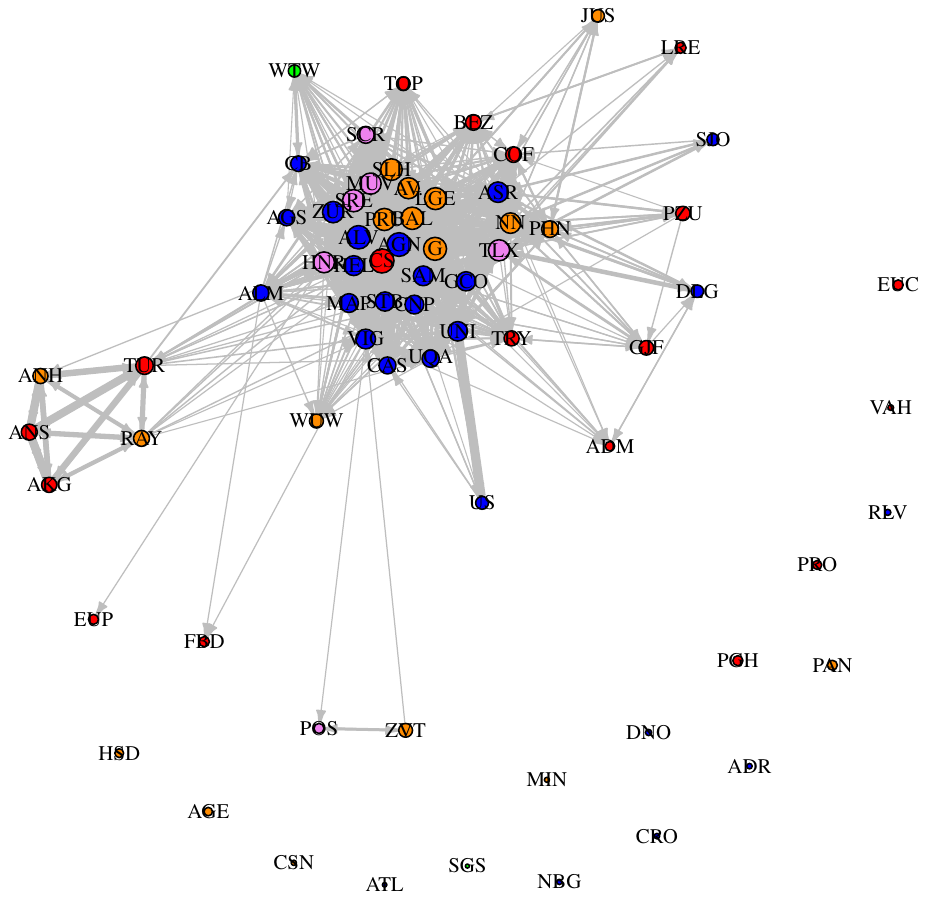}
\vspace{-2cm}
\caption{Network of insurance companies, color-coded by subsector—insurance brokers (green), life and health (orange), multiline (blue), property and casualty (red), and reinsurance (violet)—based on the GFEVD model estimated on the CAViaR time series.}
\label{fig:CompaniesNetwork_CAViaR}
\end{figure}

\begin{figure}[hbt!]
\hspace{-5.5cm}
\includegraphics[scale=0.8]{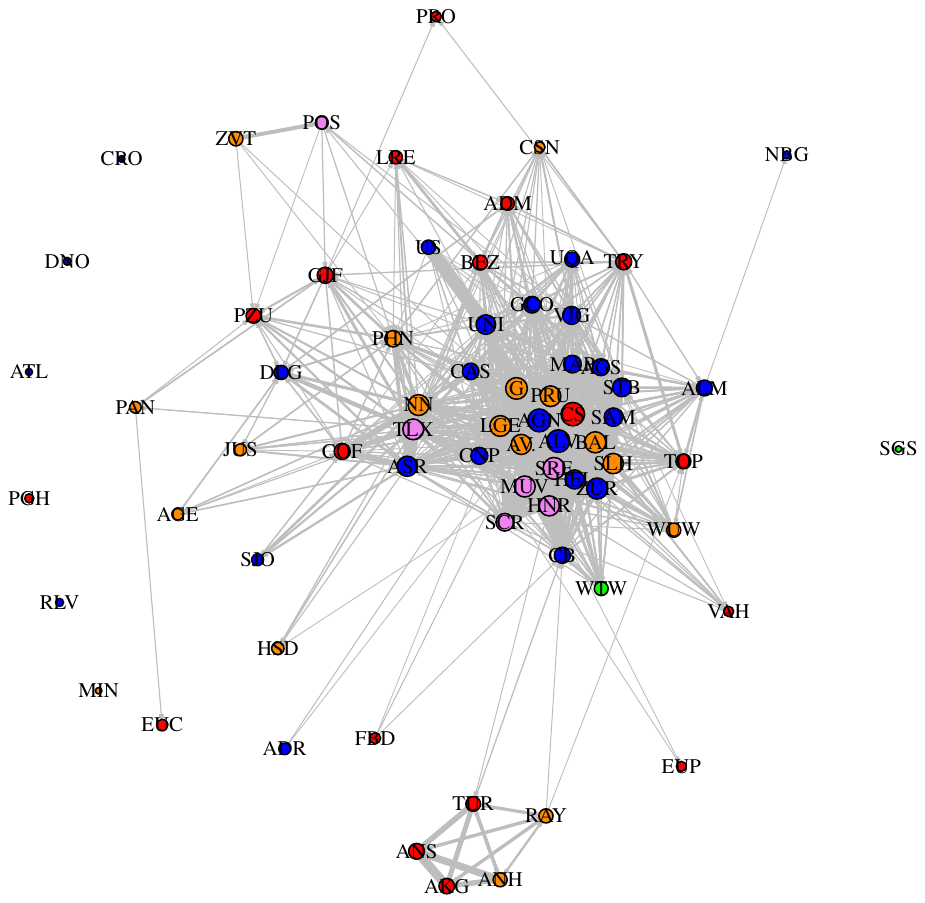}
\vspace{-2cm}
\caption{Network of insurance companies, color-coded by subsector—insurance brokers (green), life and health (orange), multiline (blue), property and casualty (red), and reinsurance (violet)—based on the GFEVD model estimated on the CARES time series.}
\label{fig:CompaniesNetwork_CARES}
\end{figure}

We construct four directed and weighted networks from the GFEVD estimation, each corresponding to a different variable: log-return (Figure \ref{fig:CompaniesNetwork_LogRet}), conditional log-volatility (Figure \ref{fig:CompaniesNetwork_LogVol}), CAViaR (Figure \ref{fig:CompaniesNetwork_CAViaR}), and CARES (Figure \ref{fig:CompaniesNetwork_CARES}). Each network consists of 70 nodes, representing the selected insurance companies, which are visually distinguished by color according to their respective subsectors: insurance brokers (green), life and health insurers (orange), multiline insurers (blue), property and casualty insurers (red), and reinsurers (violet). The size of each node $j$ is proportional to $\tilde{\theta}_{\bullet \leftarrow j}(h)$ from Equation \eqref{eq:outgoingC}, emphasizing companies with greater systemic impact. Similarly, the width of a directed link from node $j$ to node $i$ is proportional to $\tilde{\theta}_{i,j}\left(h\right)$ from Equation \eqref{eq:normgenFEVD}. 

\begin{figure}[hbt!]
\hspace{0cm}
\includegraphics[scale=0.34]{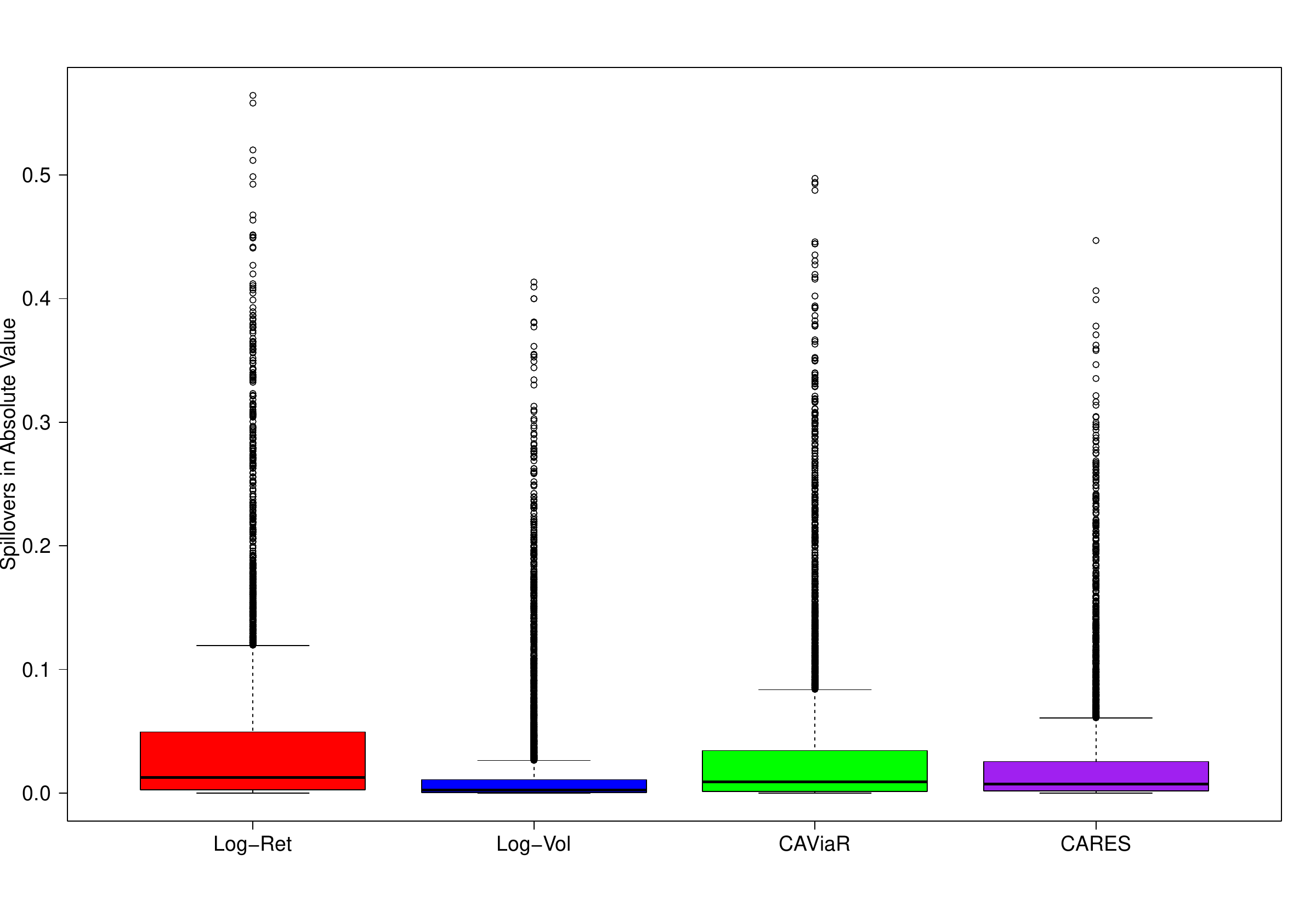}
\caption{Boxplots of the absolute values of $\tilde{\theta}_{i,j}\left(h\right)$, with $i,j=1,\ldots,70$ and $i \neq j$, resulting from the GFEVD estimation at the level of individual insurance companies on four different variables: log-returns (Log-Ret), conditional log-volatility (Log-Vol), conditional Value-at-Risk (CAViaR), and conditional Expected Shortfall (CARES). The boxes include companies above the bottom 5\% and below the top 95\% of the distribution. The horizontal lines denote the median.}
\label{fig:BoxplotSpilloversCompanies}
\end{figure}

The estimated spillovers are densely clustered around zero, as illustrated in Figure \ref{fig:BoxplotSpilloversCompanies}, which presents the distributions of $\tilde{\theta}_{i,j}\left(h\right)$ in absolute value (excluding self-loops). The four boxplots reveal that a subset of spillovers is substantially larger, forming a right tail that deviates from the bulk of the data---suggesting economically meaningful heterogeneity in spillover intensities within the network. To enhance clarity and highlight the most relevant spillovers, we exclude from Figures \ref{fig:CompaniesNetwork_LogRet}–\ref{fig:CompaniesNetwork_CARES} all links whose absolute value is below the third quartile of their respective distributions. We use the 75th percentile as the baseline threshold because it strikes a balance between retaining a sufficiently rich network and filtering out weak, noisy links, thus improving interpretability of the network structure

The networks in Figures \ref{fig:CompaniesNetwork_LogRet}–\ref{fig:CompaniesNetwork_CARES} are visualized using the \citeauthor{Fruchterman1991} \citeyearpar{Fruchterman1991} force-directed algorithm, a widely used approach for network layout optimization. This algorithm simulates a physical system where nodes repel each other while edges act as attractive forces, pulling connected nodes together. As a result, highly interconnected companies form tightly clustered groups, whereas loosely connected firms are positioned toward the periphery \citep{Bax2024}.

Figures \ref{fig:CompaniesNetwork_LogRet}\textemdash \ref{fig:CompaniesNetwork_CARES} reveal several key stylized facts. First, a well-defined core group of insurance companies consistently plays a central role across all four networks. Most of these firms belong to the multiline insurance sector, including Aegon, Ageas, Allianz, Unipol, and Zurich. Additionally, Aviva, Baloise, Generali, and Prudential emerge as central players within the life and health insurance industry. In the property and casualty sector, AXA, Coface, Topdanmark, and Tryg are prominent, with the latter three positioned at the periphery of the core cluster. The reinsurance subsector is also well represented, with Hannover Rück, Münchener Rück, and Swiss Re forming a central group. This core set of firms remains largely consistent across the networks, with only minor variations. For instance, Direct Line and Talanx appear centrally in Figure \ref{fig:CompaniesNetwork_LogRet} but become more peripheral in Figure \ref{fig:CompaniesNetwork_LogVol}, where they are still interconnected but relatively isolated from the main cluster. Notably, AXA consistently stands out as the most systemically important company across all networks, maintaining a central position.

Another important pattern is the strong clustering of companies within the same subsector, reinforcing intra-sectoral interconnectedness. Additionally, a significant ``country effect'' is evident. Turkish insurers—including Aksigorta, Anadolu Hayat Emeklilik, Anadolu Sigorta, Ray Sigorta, and Türkiye Sigorta—form a distinct, isolated cluster, exhibiting strong mutual interconnections across all four networks. A similar phenomenon is observed for the Slovenian insurers Pozavarovalnica Sava and Zavarovalnica Triglav, which are closely linked yet remain disconnected from the broader network. Although they belong to different subsectors—reinsurance and life and health insurance, respectively—their geographical proximity is consistent with a country-specific component in their interconnectedness. This country-based clustering is particularly pronounced in Figure \ref{fig:CompaniesNetwork_LogRet}.

In the construction of Figures \ref{fig:CompaniesNetwork_LogRet}\textemdash\ref{fig:CompaniesNetwork_CARES}, we exclude all links with absolute values below the third quartile of their respective distributions. To assess the robustness of this choice, and in particular the most critical aspect concerning the intersection of the central communities, we re-estimated the networks using alternative thresholds at the 60-th and 90-th percentiles, in addition to the 75-th percentile baseline. This is a demanding exercise, since the range from the 60-th to the 90-th percentile spans nearly half of the distribution's support. We find that our estimates are not only robust to the choice of threshold, but also display a clear continuity across specifications: a stable core of systemically central insurers\textemdash Aviva, Legal \& General, and Prudential\textemdash consistently emerges, while the surrounding set of firms evolves smoothly as the cutoff is relaxed or tightened. As expected, the intersection becomes more inclusive and denser when the threshold is set at the 60-th percentile, admitting a larger set of firms, namely Allianz, Baloise, Chubb, Hannover Rueck, Helvetia, Muenchener Rueckversicherungs-Gesellschaft, Swiss Life, Swiss Re, Vaudoise Assurances, Willis Towers Watson, and Zurich, in addition to those identified using the 75-th percentile. At the more restrictive 90-th percentile only the most central ``hard core'' of firms remains (Aviva, Legal \& General, and Prudential). This pattern is economically intuitive and confirms that the identification of key systemic insurers\textemdash and in particular the intersection of the central communities\textemdash is robust to the choice of threshold. These results are described in Table \ref{tab:robust_different_threshold} in the Online Appendix.

The identification of central insurance companies becomes even more precise through the application of the Louvain community detection algorithm \citep[see, among others,][]{Palowitch2019,Hegde2025}. To implement this approach, we first convert each network in Figures \ref{fig:CompaniesNetwork_LogRet}\textemdash \ref{fig:CompaniesNetwork_CARES} from directed to undirected, meaning that link direction is no longer considered. In the resulting undirected networks, a link between nodes $i$ and $j$ exists if, in the original directed network, there is a link from $i$ to $j$, from $j$ to $i$, or both. The weight of each undirected link is then computed as the sum $\left(\tilde{\theta}_{i \leftarrow j}(h)+\tilde{\theta}_{j \leftarrow i}(h)\right)$. We then apply the Louvain algorithm to identify communities within each network, focusing on the largest community (i.e., the one containing the most nodes). The intersection of the largest and most central communities detected across all four networks yields a core group of eight insurance companies: Aegon, Aviva, Ageas, Storebrand, Generali, AXA, Legal \& General, and Prudential. 

This intersection of central communities highlights a subset of insurers with a pronounced systemic impact, consistently emerging as key players across all networks. Notably, this group aligns closely with the list of Global Systemically Important Insurers (G-SIIs) published by the Financial Stability Board (FSB) from 2013 to 2016, when the designation of G-SIIs was officially suspended, as reported in Table \ref{tab:G-SIIs}. In this table, European insurers are highlighted in bold and italic, while non-European firms---American International Group, MetLife, Prudential Financial, Inc., and Ping An Insurance (Group) Company of China Ltd.---are not present in our sample of insurance companies in Eastern and Western Europe. An advantage of our methodology is that it relies solely on publicly available stock market data, whereas the FSB's classification is based on a broad range of proprietary indicators, including firm-specific exposures, cross-jurisdictional activities, and off-balance-sheet positions. Despite these methodological differences, our results show substantial overlap with the FSB's multi-criteria assessment, suggesting that the approach can provide a useful market-based screening tool for systemic relevance, rather than a substitute for balance-sheet-based supervisory assessments. These results complement those in \cite{chen2020tail}, which uses a tail risk network based on CoVaR to identify systemically relevant insurance companies.

\begin{table}[htbp]
  \centering
  \begin{footnotesize}
    \resizebox{\textwidth}{!}{ 
    \begin{tabular}{llll}
    \toprule
    \multicolumn{1}{c}{2013} & \multicolumn{1}{c}{2014} & \multicolumn{1}{c}{2015} & \multicolumn{1}{c}{2016} \\
    \midrule
    \textit{\textbf{Allianz}} & \textit{\textbf{Allianz}} & \textit{\textbf{Aegon}} & \textit{\textbf{Aegon}} \\
    American Int. & American Int. & \textit{\textbf{Allianz}} & \textit{\textbf{Allianz}} \\
    \textit{\textbf{Generali}} & \textit{\textbf{Generali}} & American Int. & American Int. \\
    \textit{\textbf{Aviva}} & \textit{\textbf{Aviva}} & \textit{\textbf{Aviva}} & \textit{\textbf{Aviva}} \\
    \textit{\textbf{AXA}}   & \textit{\textbf{AXA}}   & \textit{\textbf{AXA}}   & \textit{\textbf{AXA}} \\
    MetLife & MetLife & MetLife & MetLife \\
    Ping An Ins. & Ping An Ins. & Ping An Ins. & Ping An Ins. \\
    Prudential F.I. &  Prudential F.I. &  Prudential F.I. &  Prudential F.I. \\
    \textit{\textbf{Prudential Plc}} & \textit{\textbf{Prudential Plc}} & \textit{\textbf{Prudential Plc}} & \textit{\textbf{Prudential Plc}} \\
    \bottomrule
    \end{tabular}%
    } 
  \caption{This table reports the list of Global Systemically Important Insurers (G-SIIs) provided by the Financial Stability Board from 2013 to 2016; European companies are written in bold and italic text.}
  \label{tab:G-SIIs}
  \end{footnotesize}
\end{table}

There are, however, some differences. For instance, Allianz, which appears in Table \ref{tab:G-SIIs}, is absent from our final intersection. However, it is worth noting that Allianz is part of three out of the four central communities identified by our method. Conversely, three companies in our intersection\textemdash Ageas, Storebrand, and Legal \& General\textemdash are not included in the FSB's list of G-SIIs. Despite these discrepancies, our findings suggest that our approach provides a useful proxy for systemic importance, successfully identifying most of the insurers classified by the FSB while relying solely on publicly available market data.

\begin{figure}[hbt!]
\hspace{0cm}
\includegraphics[scale=0.33]{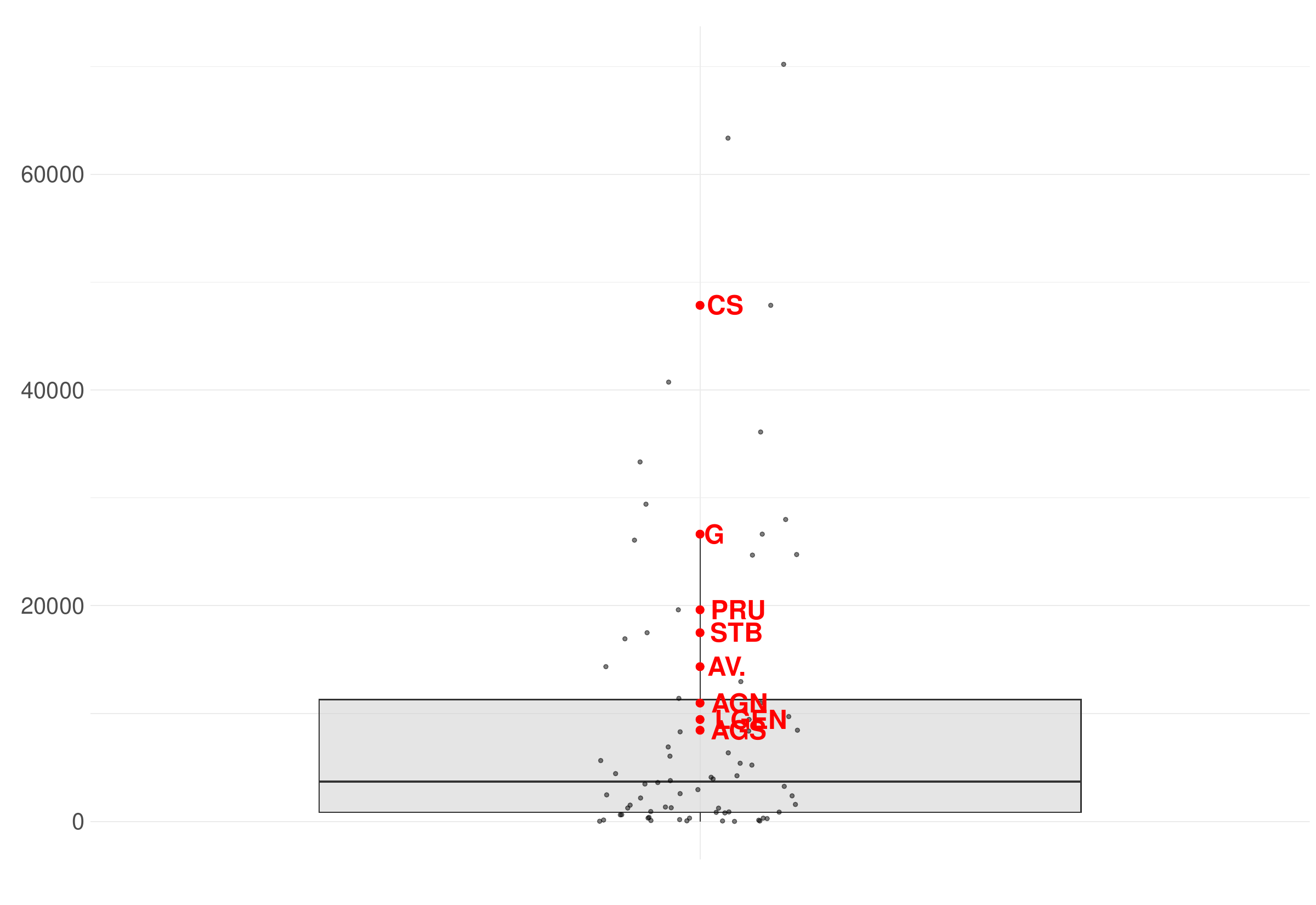}
\caption{Distribution of the average market capitalization of the 70 insurance companies included in the networks given in Figures \ref{fig:CompaniesNetwork_LogRet}\textemdash \ref{fig:CompaniesNetwork_CARES}. Market capitalization is in millions of euros. The gray box includes companies with a market capitalization above the bottom 5\% and below the top 95\% of the distribution. The horizontal line is the median.}
\label{fig:BoxplotSizeCompanies}
\end{figure}

An important question is whether the systemic impact of the analyzed insurance companies is purely driven by their size. To investigate this, we use average market capitalization over the analyzed period as a proxy for firm size. Figure \ref{fig:BoxplotSizeCompanies} presents a boxplot of these market capitalizations, highlighting in red the eight most systemically important companies identified through our intersection. While these firms are generally large (positioned above the median size), size alone does not fully explain their systemic relevance. Notably, although AXA is the most central institution in our analysis, other companies with comparable or even larger market capitalization do not necessarily exhibit the same systemic importance in the networks shown in Figures \ref{fig:CompaniesNetwork_LogRet}\textemdash \ref{fig:CompaniesNetwork_CARES}. 

This finding underscores that being large is not sufficient to be systemically central in our network-based measures. Instead, measured systemic importance is more closely associated with interconnectedness and the structure of financial dependencies within the insurance system. Firms with extensive linkages and central positions in the network can exhibit a greater systemic footprint in our spillover measures, regardless of their absolute size.

\subsection{Robustness checks}\label{sec:robustness}

In this section, we conduct a series of robustness checks to evaluate the stability of our results with respect to key parametric and empirical choices made in the previous sections. We focus on the individual insurance company level, which presents the greatest challenge among the three aggregation levels\textemdash financial markets, insurance subsectors, and individual firms\textemdash due to the significantly larger number of variables, potentially increasing result variability. To address the curse of dimensionality, we apply the post-LASSO method for estimating the VAR parameters, as discussed in Section \ref{sec:fevd}.  

Maintaining the same setup as in Section \ref{sec:companies}, we re-estimate the GFEVD model on the 70 insurance companies for the balanced sample. For simplicity, this section focuses on the GFEVD model estimated on log-return time series.\footnote{Estimates obtained from conditional log-volatility, CAViaR, and CARES time series, as well as those derived from aggregated financial markets and insurance subsectors, are available upon request.}

\begin{figure}[hbt!]
\hspace{0cm}
\includegraphics[scale=0.33]{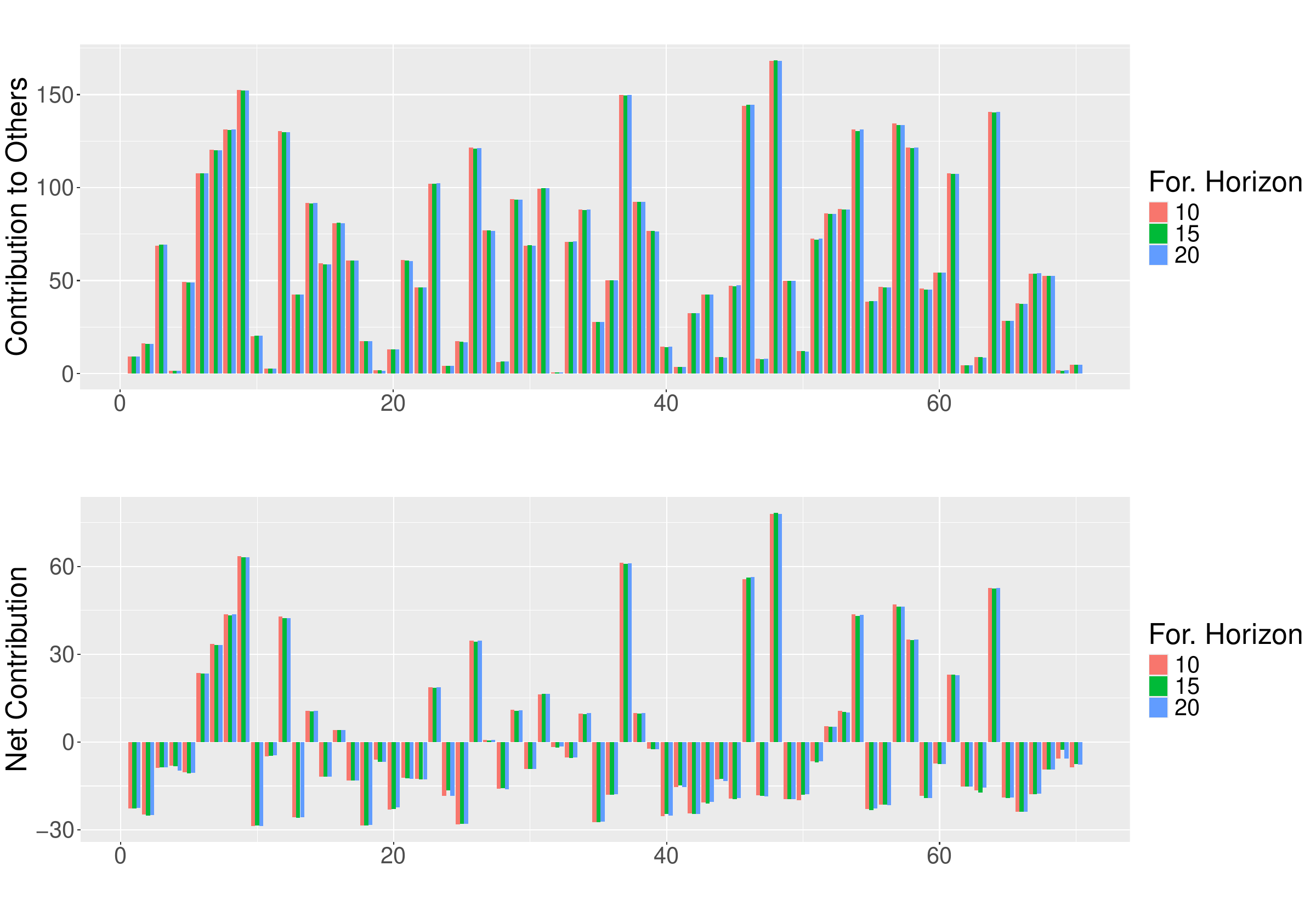}
\caption{Contributions to others and net contributions of the 70 insurance companies identified with a value of ``1'' in the fourth column of Table \ref{tab:infocompanies} in \ref{app:tablesfigures}. The results are derived from the GFEVD model estimated on daily log-return time series for different forecast horizons ($h$): 10, 15, and 20, with the underlying VAR model set to $p=1$. The VAR parameters are estimated using the post-LASSO method.}
\label{fig:Robustness_ForecastHorizonRet}
\end{figure}

We begin by assessing the impact of varying the forecast horizon $h$ in Equation \eqref{eq:genFEVD} on our estimates. While the previous sections adopted $h=10$, we now extend the analysis to $h=15$ and $h=20$, keeping the lag order fixed at $p=1$. The top panel of Figure \ref{fig:Robustness_ForecastHorizonRet} displays the values of $\tilde{\theta}_{\bullet \leftarrow j}(h)$, as defined in Equation \eqref{eq:outgoingC}, which measures the contribution of company $j$ to others, for $j=1,\ldots,70$ and $h=10,15,20$. The bottom panel of Figure \ref{fig:Robustness_ForecastHorizonRet} presents the net contributions, computed as the difference between contributions to and from others, for the same set of companies.

The results remain remarkably stable across different values of $h$, with only minor variations among the three cases. This consistency reinforces the robustness of our approach, indicating that our methodology effectively isolates and quantifies systemic spillovers without being overly sensitive to the choice of forecast horizon.

\begin{figure}[hbt!]
\hspace{0cm}
\includegraphics[scale=0.33]{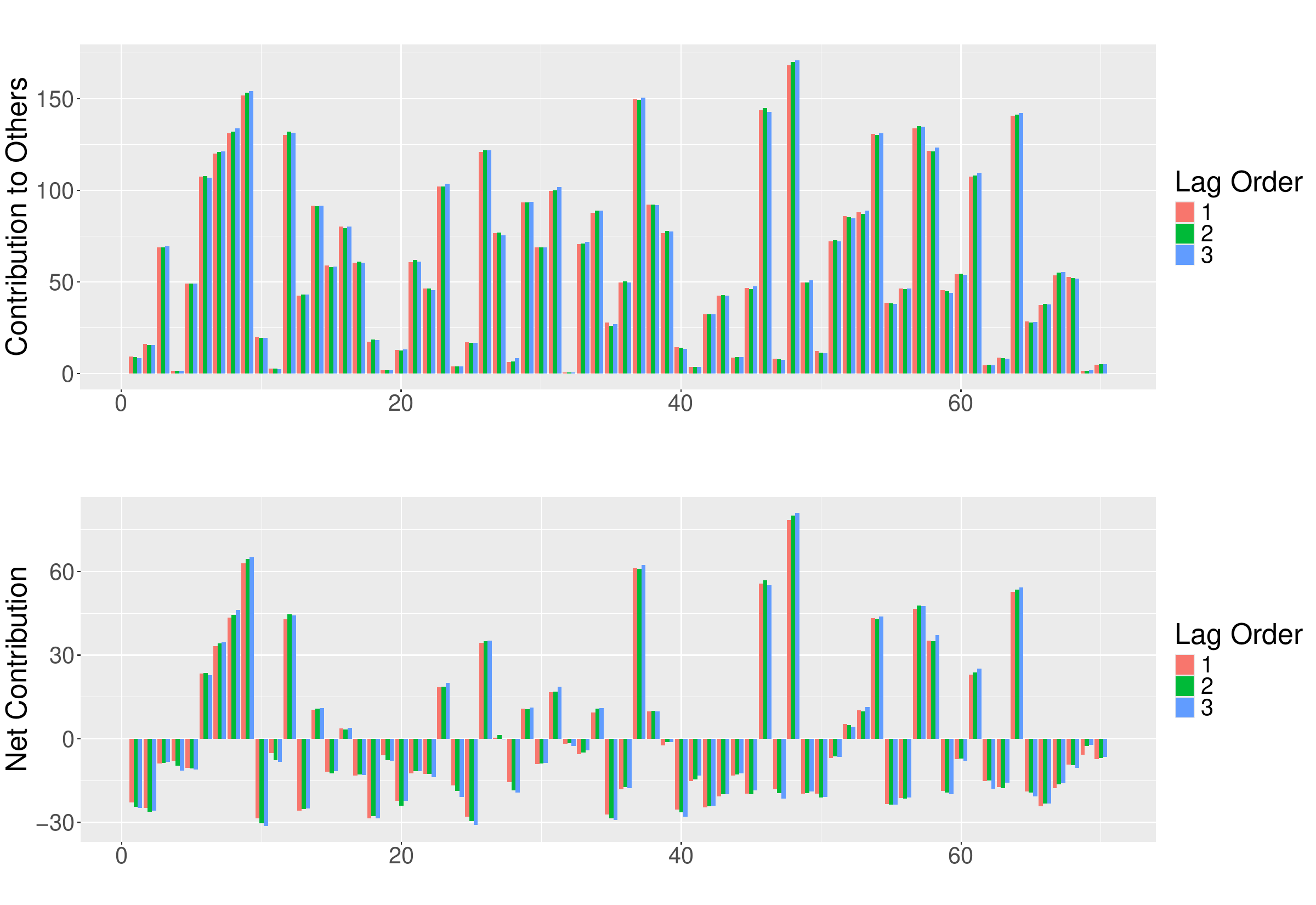}
\caption{Contributions to others and net contributions of the 70 insurance companies identified with a value of ``1'' in the fourth column of Table \ref{tab:infocompanies} in \ref{app:tablesfigures}. The results are derived from the GFEVD model with $h=10$, estimated on daily log-return time series for different lag orders ($p$) in the underlying VAR model: 1, 2, and 3. The VAR parameters are estimated using the post-LASSO method.}
\label{fig:Robustness_LagOrderRet}
\end{figure}

We draw similar conclusions regarding the lag order. In particular, Figure \ref{fig:Robustness_LagOrderRet} presents the estimates obtained from the GFEVD model with a forecast horizon of $h=10$, based on VAR parameters estimated using lag orders $p=1$, $p=2$, and $p=3$. The estimates remain nearly identical across the different values of $p$.  

\begin{figure}[hbt!]
\hspace{0cm}
\includegraphics[scale=0.33]{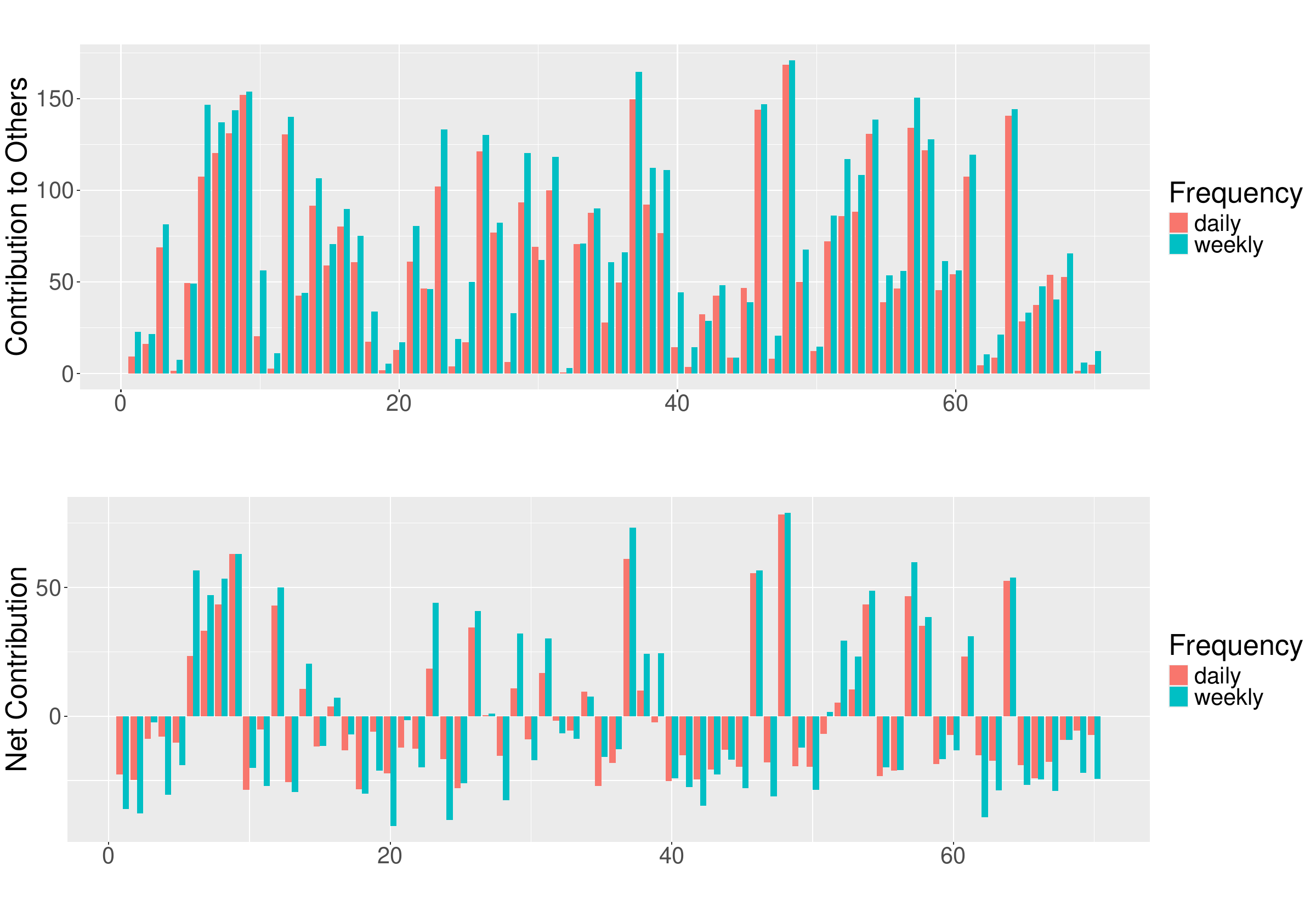}
\caption{Contributions to others and net contributions of the 70 insurance companies marked with the value ``1'' in the fourth column of Table \ref{tab:infocompanies} given in  \ref{app:tablesfigures}. The results are obtained from the GFEVD model ($h=10$) estimated on the weekly log-return time series, setting $p=1$ in the underlying VAR model. The VAR parameters are estimated using the post-LASSO method.    
}
\label{fig:RobustnessCompanies_WeeklyRet}
\end{figure}

Our final robustness check examines the impact of data frequency. While our empirical analysis is based on daily log-returns, we now assess how the results change when the frequency is reduced to weekly log-returns. Figure \ref{fig:RobustnessCompanies_WeeklyRet} compares the contributions to others and net contributions of the selected insurance companies, obtained from estimating the GFEVD model using both daily and weekly log-returns. 

Compared to Figures \ref{fig:Robustness_ForecastHorizonRet} and \ref{fig:Robustness_LagOrderRet}, the differences between the bars for each company in Figure \ref{fig:RobustnessCompanies_WeeklyRet} are relatively larger. However, these variations remain minimal, reinforcing the robustness of our methodology across different data frequencies.

\section{Conclusions}\label{sec:conclusions}

This paper studies systemic-risk connectedness in the European insurance sector at three levels of granularity: across major financial markets, across insurance subsectors, and across individual insurance companies. Using a common connectedness framework applied to returns, volatility, VaR, and ES, we document that insurers are an important component of the European financial system and that their spillovers are strongly time-varying, especially during stress episodes.

In the aggregate time-series analysis, spillovers from insurance are more strongly associated with the term spread, sovereign spread, and funding stress, while a broad global risk-appetite proxy is less informative. In addition, a firm-level panel exercise on individual insurers shows that the association between sovereign stress and insurer-to-bank spillovers varies with domestic sovereign-bond home bias: sovereign stress is associated with lower insurer-to-bank spillovers for insurers in countries with low home bias, but this relationship weakens and could reverse in countries with very large home bias. We interpret these findings as suggestive evidence consistent with a balance-sheet channel in the European institutional setting, while emphasizing that the evidence is reduced-form and not causal.

At the firm level, the network analysis identifies a stable core of highly central insurers across multiple risk indicators, and the overlap with the historical G-SII list provides an economically meaningful external benchmark for the connectedness measures. More broadly, our findings support the use of spillover metrics as monitoring tools that can help supervisors prioritize deeper analysis, for example by identifying central firms, vulnerable subsectors, or periods of heightened tail-risk transmission.

At the same time, these measures are market-price-based and reduced-form, and therefore do not separately identify structural channels or causal policy effects. For this reason, we view our results as disciplined descriptive evidence and mechanism-consistent correlations, to be complemented by balance-sheet and institutional analysis in policy applications. This is particularly relevant in Europe, where ongoing discussions on financial stability continue to address the role of sovereign exposures, the possible development of a common safe asset, and the prudential treatment of government bonds.

Furthermore, the recent Draghi report underscores the importance of the Capital Markets Union while highlighting the challenges posed by fragmented European capital markets and heterogeneous supervisory competences across member countries. In light of our findings, a stronger European supervisory capacity to monitor and assess systemic spillover risks in the insurance sector appears increasingly important. Such a function could be strengthened within the existing European Systemic Risk Board framework or, more ambitiously, assigned to a dedicated European insurance supervisory authority for the largest and most systemically significant institutions, broadly analogous to the centralized model already in place for banks.


\newpage


\bibliographystyle{elsarticle-harv} 
\bibliography{RefSystemicRisk.bib}

\clearpage


\appendix

\setcounter{section}{0} \renewcommand{\thesection}{\Alph{section}} \setcounter{table}{0} \renewcommand{\thetable}{A.\arabic{table}} \setcounter{figure}{0} \renewcommand{\thefigure}{A.\arabic{figure}}
\setcounter{page}{1}

\section*{Appendix}

\section{Performance and risk indicators}
\label{sec:metindicators}

We estimate the GFEVD model described in Section \ref{sec:fevd} using four indicators of performance and risk based on equity prices. 

The first indicator is the log-return, a standard measure of market performance.  Let $r_{j,t}$ be the log-return yielded by a given financial entity at time $t$, we estimate the GFEVD model by setting $\bfx_t=\left[x_{1,t}=r_{1,t} \cdots x_{n,t}=r_{n,t}\right]^\prime$.  

The second indicator is the conditional log-volatility, a standard measure of risk in financial markets, that we model using the Generalized AutoRegressive Conditional Heteroskedasticity (GARCH) model \citep{Bollerslev1986}. We employ the following AutoRegressive Moving-Average (ARMA) process for the mean equation:
\begin{equation}\label{eq:armamean}
r_{j,t}=\phi_{j,0} + \sum_{i=1}^k \phi_{j,i} r_{j,t-i} + \sum_{i=1}^q \xi_{j,i} a_{j,t-i},
\end{equation}
where $a_{j,t}$ is the innovation or shock of $r_{j,t}$ at time $t$ \citep{Tsay2010},  defined as:
\begin{equation}
a_{j,t} = \widetilde{\sigma}_{j,t} \widetilde{\epsilon}_{j,t},
\end{equation}
with $\{ \widetilde{\epsilon}_{j,t} \}$ being a sequence of independent and identically distributed random variables with mean zero and variance one. 

The standard variance equation \citep{Bollerslev1986} has the following specification: 
\begin{equation}\label{eq:garchvarequation}
\widetilde{\sigma}^2_{j,t} = \omega_0 + \sum_{i=1}^m \omega_{j,i} a^2_{j,t-i} + \sum_{i=1}^s \nu_{j,i} \widetilde{\sigma}^2_{j,t-i},
\end{equation}
where $\omega_{j,i}$ and $\nu_{j,i}$ are refereed to as ARCH and GARCH parameters, respectively \citep{Tsay2010}.

Starting from the specification in Equation  \eqref{eq:garchvarequation} introduced by \cite{Bollerslev1986},  alternative variance equations have been proposed in the literature. In our study, we consider the EGARCH \citep{Nelson1991}, and the GJR GARCH \citep{GLOSTEN1993} models, in addition to the standard specification in \eqref{eq:garchvarequation}.  Moreover, we adopt six different distributions for the $\widetilde{\epsilon}_{j,t}$ random variable: the Gaussian,  Student-t,  and Generalized Error distributions,  and their respective skew variants based on the transformations described in \cite{Fernandez1998} and \cite{Ferreira2006}. We test the AR and MA orders\textemdash i.e.  $k$ and $q$ in Equation \eqref{eq:armamean}\textemdash as well as the orders of the ARCH and GARCH effects\textemdash i.e.  $m$ and $s$ in Equation \eqref{eq:garchvarequation}, and in the alternative EGARCH and GJR GARCH variance equations\textemdash from zero to two. We combine all the aforementioned alternative choices, and select, for each individual time series, the best GARCH model based on the BIC.  From the best GARCH specification, we then extract the time series of the estimated conditional volatility, denoted as $\widetilde{\sigma}^\star_{j,t}$, and estimate the GFEVD model on 
$\bfx_t=\left[x_{1,t}=  \ln\left(\widetilde{\sigma}^\star_{1,t}\right) \cdots  x_{n,t}=  \ln\left(\widetilde{\sigma}^\star_{n,t}\right)   \right]^\prime$. 

Next, we take the value-at-risk (VaR), a standard measure of tail risk, to build the third indicator of performance and risk.  Specifically,  we estimate the Conditional AutoRegressive VaR (CAViaR) model introduced by \cite{Engle2004} to obtain a time series of time-varying VaR values.  In contrast to other approaches that model entire return distributions to extract quantiles, the CAViaR model directly models quantiles on the basis of AR specifications. The underlying parameters are estimated through quantile regression \citep{Koenker1978}.  Let $f_t \left( \bfgamma  \right) \equiv f_t\left( \bfz_{t-1},\bfgamma_\tau \right)$ be the latent quantile at time $t$ to estimate from a given return time series, where $\tau$ is the probability associated to VaR, and $\bfz_t$ is a vector of time $t$ observable variables.\footnote{We set $\tau=0.05$ in our empirical analysis.}  The generic CAViaR specification takes the following form:
\begin{equation}\label{eq:CAViaR}
f_t\left( \bfgamma \right) = \gamma_0 + \sum_{i=1}^q \gamma_i f_{t-i}\left( \bfgamma  \right) + \sum_{j=1}^r \gamma_j \ell (\bfz_{t-j}).
\end{equation}

The AR terms $\gamma_1 f_{t-1}\left( \bfgamma  \right), \ldots, \gamma_q f_{t-q}\left( \bfgamma  \right)$ in Equation \eqref{eq:CAViaR} ensure that the latent quantiles smoothly change over time.  The role of $\ell (\bfz_{t-j})$ is to link
$f_t\left( \bfgamma \right)$ to observable variables that belong to the information set \citep{Engle2004}.  Among the alternative specifications proposed by \cite{Engle2004}, we adopt the asymmetric slope function to estimate the CAViaR of the $j$-th return time series:
\begin{equation}\label{eq:asCAViaR}
f_{t}\left( \bfgamma_j \right) = \gamma_{j,1} + \gamma_{j,2} f_{t-1}\left( \bfgamma_j  \right) + \gamma_{j,3} \left( r_{j,t-1} \right)^+ + \gamma_{j,4} \left( r_{j,t-1} \right)^-,
\end{equation}
where $\left( r_{j,t-1} \right)^+ = \max \left(r_{j,t-1},0   \right)$, and $\left( r_{j,t-1} \right)^- = -\min \left(r_{j,t-1},0   \right)$.

We use the asymmetric slope specification to take into account potential asymmetric effects of past returns on VaR.  By doing so, we differentiate between the effects of positive and negative lagged returns on the estimated quantile.  This specification is particularly useful in financial frameworks where extreme negative returns might influence risk estimates more strongly than positive returns \citep{Engle2004}.  After obtaining the CAViaR time series from Equation \eqref{eq:asCAViaR} for $j=1,\ldots,n$, we then estimate the GFEVD model by setting $\bfx_t=\left[x_{1,t}=  f_{t}\left( \bfgamma_1 \right) \cdots  x_{n,t}=  f_{t}\left( \bfgamma_n \right)   \right]^\prime$.  

Furthermore, we consider the expected shortfall (ES) as a fourth measure of performance and risk.  In contrast to VaR,  that does not necessarily satisfy the subadditivity property,  ES is a coherent measure of risk \citep{Artzner1999}.  Moreover,  VaR only gives the threshold loss level at a specific confidence level, but ignores what happens beyond this point, while ES captures the expected loss at a given confidence level. The better properties of ES are reflected in the recent changes in financial regulation. In fact,  Basel III guidelines recommend ES over VaR for calculating market risk in financial institutions. 

In general,  ES is defined as the expected return of a given asset conditional on the fact that such return is less than its VaR at a given confidence level $\tau$.  We obtain a time-varying ES by adopting the approach proposed by \cite{Taylor2007}, that builds on expectiles.  In particular,  the population expectile of the random variable $R_j$ at level $\psi$ is defined as the quantity $\delta \left(R_j,\psi \right)$ that minimizes the following expected loss:
\begin{equation}\label{popexp}
\mathbb{E}\left[ \left| \psi - \mathbb{I}_{\left\{R_j < \delta \left(R_j,\psi \right)  \right\}}  \right| \left(R_j- \delta \left(R_j,\psi \right) \right)^2 \right],
\end{equation} 
where $\psi \in (0,1)$, and $\mathbb{I}_{\{\cdot\}}$ is an indicator function which takes the value of one if the condition into braces is true, and the value of zero otherwise. 

\cite{Taylor2007} showed that the ES of $R_j$ at the confidence level $\tau$, denoted as $\text{ES}\left(R_j,\tau  \right)$, is a function of $\delta \left(R_j,\psi \right)$ through the following relationship:
\begin{equation}\label{exp1}
\text{ES}\left(R_j,\tau  \right) = \left(1+\frac{\psi}{(1-2 \psi)\tau}    \right) \delta \left(R_j,\psi \right) - \frac{\psi}{(1-2 \psi)\tau} \mathbb{E}[R_j], 
\end{equation}
where $\mathbb{E}[R_j]$ is the expected value of $R_j$.

The relationship given in Equation \eqref{exp1} is defined for scalar expectile and ES.  However,  \cite{Taylor2007} showed that such relationship holds even if expectiles are conditional on a set of explanatory variables, for instance, through the following Conditional AutoRegressive Expectile (CARE) model: 
\begin{equation}\label{care}
\delta \left(r_{j,t},\psi \right) = \eta_{0,j, \psi} + \eta_{1,j,\psi} \delta \left(r_{j,t-1},\psi \right)+ \eta_{2,j,\psi} |r_{j,t-1}|.
\end{equation}

Therefore, similar to CAViaR,  the CARE model captures the persistence and dynamics of expectiles over time. In empirical applications, \cite{Taylor2007} showed that the time-varying ES at the probability level $\tau$ coincides with the $\psi_j^\star$-th expectile, denoted as $\delta \left(r_{j,t},\psi_j^\star \right)$, where $\psi_j^\star$ is the value of the expectile level $\psi$ given in Equation \eqref{care} that makes the percentage of in-sample violations equal to $\tau$:
\begin{equation}\label{eq:percisviol}
\left(T-1\right)^{-1} \sum_{t=2}^{T} \mathbb{I}_{\left\{r_{j,t} < \delta \left(r_{j,t},\psi_j^\star \right)  \right\}}=\tau.
\end{equation}

This rule leads to the Conditional AutoRegressive Expected Shortfall (CARES) model of \cite{Taylor2007}.\footnote{Following \cite{Bonaccolto2022}, in the empirical analysis we iteratively estimate the CARE model in Equation \eqref{care} using a dense sequence of $\psi$ values with starting point $10,000^{-1}$. We then select the $\psi$ value that produces the minimum distance between the percentage of in-sample violations and $\tau$, setting it as $\psi_j^\star$.} We stress the fact that $\psi_j^\star$ and $\psi_i^\star$ do not necessarily take the same values, for $j,i=1,\ldots,n$ and $j \neq i$. However, $\delta \left(r_{j,t},\psi_j^\star \right)$ and $\delta \left(r_{i,t},\psi_i^\star \right)$ are both ESs at the probability level $\tau$. Therefore, we finally estimate the GFEVD model on the time-varying ESs, by setting $\bfx_t=\left[x_{1,t}=  \delta \left(r_{1,t},\psi_1^\star \right) \cdots  x_{n,t}=  \delta \left(r_{n,t},\psi_n^\star \right)   \right]^\prime$.

\section{Macro-financial correlates of insurers' tail spillovers}\label{sec:ts_regs}

The previous subsection documents substantial time variation in insurers' directional tail spillovers to other
financial markets. We next relate this variation to a small set of macro-financial state variables that proxy for
interest-rate (duration) risk, sovereign stress, funding conditions, and global risk appetite. Specifically, we
focus on the expected shortfall (ES) spillover from aggregate insurance to (i) all asset markets,
(ii) equities (excluding financials), (iii) banks, and (iv) government bonds. The spillover series are constructed
from rolling-window estimates sampled at weekly frequency using end-of-week values.]

For each destination market \(j \in \{\text{All}, \text{Equity}, \text{Banks}, \text{Bonds}\}\), we estimate the
following reduced-form specification in levels:
\begin{equation}\label{eq:ts_levels_nolag}
\begin{split}
C^{ES}_{I\rightarrow j,t}
&=
\alpha_j
+
\beta_{1j}\big(y^{10y,DE}_t-y^{1y,DE}_t\big)
+
\beta_{2j}\big(y^{10y,IT}_t-y^{10y,DE}_t\big) \\
&\quad
+
\beta_{3j}\big(\text{Euribor}^{3m}_t-\text{OIS}^{3m}_t\big)
+
\beta_{4j}\Delta \mathrm{VIX}_t
+
u_{j,t}.
\end{split}
\end{equation}

where \(C^{ES}_{I\rightarrow j,t}\) is the weekly directional spillover from aggregate insurance to market \(j\).
The regressors capture four broad channels: (i) the German term spread (10-year minus 1-year yield), as a proxy for
the duration/rate channel; (ii) the Italy--Germany 10-year sovereign spread, capturing euro-area sovereign stress;
(iii) the 3-month Euribor minus 3-month OIS spread, capturing funding stress; and (iv) the weekly
change in the VIX, capturing global risk appetite. We emphasize that Eq.~\eqref{eq:ts_levels_nolag} is intended to
capture \emph{contemporaneous correlations} rather than causal effects. Using weekly data mitigates high-frequency
microstructure noise, and we intentionally avoid including a lagged dependent variable, which can mechanically
absorb much of the variation in persistent spillover series and shift the interpretation toward a dynamic
prediction exercise.

Table~\ref{tab:ins_spillovers_econ_conditions} summarizes the results. Three findings stand out.
First, at the system-wide level (insurance spillovers to all assets), the term spread and funding stress are
strongly positively associated with insurers' ES spillovers, and the sovereign spread becomes
economically and statistically important when spillovers are estimated using the longer rolling window
(\(W=504\)). Second, for spillovers from insurance to equity markets, the dominant correlate is the term
spread: a steeper yield curve is associated with higher insurance-to-equity tail spillovers, consistent with a
discount-rate channel whereby higher long-horizon rates raise discount factors and depress equity valuations.
Third, spillovers from insurance to government bonds are positively related to the term spread, the
sovereign spread, and funding stress, consistent with the importance of interest-rate and
sovereign-risk channels for insurers' bond-heavy balance sheets.

The spillovers from insurance to banks also load positively on the term spread and funding stress, while the
sovereign spread enters with a negative coefficient. This negative association is consistent with a
``patient capital'' interpretation. Insurers are typically long-horizon investors and may absorb temporary
increases in sovereign risk by holding long-dated bonds through episodes of spread widening, thereby attenuating
the directional spillover from insurers to banks. By contrast, a steepening of the yield curve captures
interest-rate and duration risk that mechanically reduces the market value of insurers' bond portfolios, tightens
solvency constraints, and is associated with higher insurance-to-bank tail spillovers. We stress that this
interpretation is suggestive: during sovereign stress episodes, tail-risk spillovers may reallocate between banks
and non-bank financial institutions, and the directional variance shares can shift accordingly as the dominant
source of extreme shocks changes over time.

Across specifications, the weekly change in the VIX is not statistically significant, suggesting that euro-area
rate, sovereign, and funding conditions are more informative for the time variation in these directional ES
spillovers than a broad global risk-appetite proxy.

\begin{table}[htbp]
\begin{footnotesize}
  \centering
  \begin{tabular}{lC{2.5cm}C{2.5cm}C{2.4cm}C{2.0cm}}
    \toprule
     & All assets & Equities ex-fin. & Banks & Govt bonds \\
    \midrule
    \multicolumn{5}{c}{Panel A: Rolling window \(W=250\) trading days} \\
    \midrule
    Term Spread      & 4.488*** & 1.300*** & 2.397*** & 0.791** \\
                    & (0.746)  & (0.334)  & (0.389)  & (0.289) \\
    Sovereign Spread & 0.905    & 0.395    & -1.073** & 1.584*** \\
                    & (0.716)  & (0.312)  & (0.376)  & (0.322) \\
    Funding Stress   & 9.351*** & 0.106    & 4.738*** & 4.507*** \\
                    & (1.795)  & (0.777)  & (0.981)  & (0.983) \\
    Risk Appetite    & -0.001   & -0.002   & -0.001   & 0.002 \\
                    & (0.006)  & (0.003)  & (0.003)  & (0.003) \\
    Constant         & 53.914*** & 23.122*** & 26.228*** & 4.564*** \\
                    & (1.362)   & (0.821)   & (0.658)   & (0.591) \\
    \midrule
    \multicolumn{5}{c}{Panel B: Rolling window \(W=504\) trading days} \\
    \midrule
    Term Spread      & 4.335*** & 1.172*** & 1.963*** & 1.200*** \\
                    & (0.531)  & (0.280)  & (0.303)  & (0.214) \\
    Sovereign Spread & 2.090*** & 0.453    & -0.581*  & 2.219*** \\
                    & (0.588)  & (0.283)  & (0.265)  & (0.284) \\
    Funding Stress   & 5.666*** & -0.837   & 3.581*** & 2.922*** \\
                    & (1.272)  & (0.758)  & (0.772)  & (0.720) \\
    Risk Appetite    & 0.006    & 0.001    & 0.003    & 0.003 \\
                    & (0.004)  & (0.003)  & (0.002)  & (0.002) \\
    Constant         & 55.039*** & 24.356*** & 27.276*** & 3.407*** \\
                    & (1.013)   & (0.699)   & (0.484)   & (0.435) \\
    \bottomrule
  \end{tabular}%

  \caption{This table reports OLS regressions of directional expected shortfall (ES) spillovers from the aggregate insurance
  sector to each destination market. The dependent variables are end-of-week values of daily ES connectedness
  estimates obtained from a rolling-window VAR/GFEVD framework with window length \(W \in \{250,504\}\) trading days.
  Term Spread is the German term spread (10-year minus 1-year government yields). Sovereign Spread is the
  Italy--Germany 10-year government yield spread. Funding Stress is the 3-month Euribor minus 3-month OIS spread.
  Risk Appetite is the weekly change in the VIX. HAC standard errors are in parentheses. Significance stars denote the
  1\%, 5\%, and 10\% levels.}
  \label{tab:ins_spillovers_econ_conditions}
\end{footnotesize}
\end{table}

\section{Additional tables and figures}\label{app:tablesfigures}

\begin{footnotesize}
\begin{longtable}{llllc}
	\caption{List of insurance companies}
	\label{tab:infocompanies}\\
	\hline
	\toprule
	Label & Company & Country & Subsector & Network \\ 
	\endfirsthead
	\midrule
	Label & Company & Country & Subsector & Network \\ 
	\midrule
	\endhead
	\hline \multicolumn{5}{c}{\textit{Continued on next page}} \\
	\endfoot
	\endlastfoot
	\midrule
	%
    ADM   & Admiral & GB    & Pro.Cas. & 1 \\
    ADRS  & Adris Grupa & HR    & Mul.Lin. & 1 \\
    AGESA & Agesa Hayat Ve Emeklilik & TU    & Lif.Hea. & 1 \\
    AGN   & Aegon & NE    & Mul.Lin. & 1 \\
    AGS   & Ageas & BE    & Mul.Lin. & 1 \\
    AKGRT & Aksigorta & TU    & Pro.Cas. & 1 \\
    ALMB  & Alm Brand & DE    & Mul.Lin. & 1 \\
    ALPTR & Patris Investimentos SGPS & PO    & Lif.Hea. & 0 \\
    ALV   & Allianz & GE    & Mul.Lin. & 1 \\
    AMSO  & AMS Osiguranje & RR    & Others & 0 \\
    ANHYT & Anadolu Hayat Emeklilik & TU    & Lif.Hea. & 1 \\
    ANSGR & Anadolu Anonim Turk Sigorta & TU    & Pro.Cas. & 1 \\
    ARMBRK2 & UNIQA Osiguranje & BY    & Mul.Lin. & 0 \\
    ASRNL & ASR Nederland & NE    & Mul.Lin. & 1 \\
    ASSI  & Assiteca Assicurativo & IT    & Ins.Bro. & 0 \\
    ATL   & Atlantic Insurance & CC    & Mul.Lin. & 1 \\
    AV.   & Aviva & GB    & Lif.Hea. & 1 \\
    BALN  & Baloise Holding & SZ    & Lif.Hea. & 1 \\
    BEZ   & Beazley & GB    & Pro.Cas. & 1 \\
    BSOSRK1 & BSO   & BY    & Mul.Lin. & 0 \\
    BSRSRK2 & Bosna Reosiguranje & BY    & Reins. & 0 \\
    CASS  & Genertel & IT    & Mul.Lin. & 1 \\
    CB    & Chubb & SZ    & Mul.Lin. & 1 \\
    CBP   & Curtis Banks & GB    & Others & 0 \\
    CNP   & CNP Assurances Saca & FR    & Mul.Lin. & 1 \\
    COFA  & Coface & FR    & Pro.Cas. & 1 \\
    CONT  & Contract Ins. and Fin. Services & GR    & Lif.Hea. & 0 \\
    COS   & Cosmos & CC    & Pro.Cas. & 0 \\
    CRHVP & Credimo Holding & BE    & Lif.Hea. & 0 \\
    CROS  & Croatia Osiguranje & HR    & Mul.Lin. & 1 \\
    CS    & AXA   & FR    & Pro.Cas. & 1 \\
    CSN   & Chesnara & GB    & Lif.Hea. & 1 \\
    DEOS  & Generali Osiguranje Monten. & YV    & Pro.Cas. & 0 \\
    DFV   & DFV Deutsche Familienvers. & GE    & Pro.Cas. & 0 \\
    DLG   & Direct Line & GB    & Mul.Lin. & 1 \\
    DLLLF & Delta Lloyd & NE    & Mul.Lin. & 0 \\
    DNOS  & Dunav Osiguranje & RR    & Mul.Lin. & 1 \\
    DNREM & Dunav & RR    & Reins. & 0 \\
    DROSRA & Drina Osiguranje & BY    & Mul.Lin. & 0 \\
    DZHS  & Source Insurance & UK    & Pro.Cas. & 0 \\
    ENGR  & Energogarant OJSC & RU    & Mul.Lin. & 0 \\
    EUC   & Europejskie Centrum Odszk. & PD    & Pro.Cas. & 1 \\
    EUMK  & Euromak Broker & MC    & Lif.Hea. & 0 \\
    EUPIC & European Reliance General & GR    & Pro.Cas. & 1 \\
    FBD   & FBD   & IR    & Pro.Cas. & 1 \\
    G     & Generali & IT    & Lif.Hea. & 1 \\
    GCL   & Lifestar Holding & MB    & Others & 0 \\
    GCO   & Grupo Catalana Occidente & SP    & Mul.Lin. & 1 \\
    GJF   & Gjensidige Forsikring & NO    & Pro.Cas. & 1 \\
    GRAW  & Grawe Osiguranje & YV    & Ins.Bro. & 0 \\
    HELN  & Helvetia Holding & SZ    & Mul.Lin. & 1 \\
    HNR1  & Hannover Rueck & GE    & Reins. & 1 \\
    HSD   & Hansard Global & IO    & Lif.Hea. & 1 \\
    HUW   & Helios Underwriting & GB    & Reins. & 0 \\
    INGS  & Ingosstrakh & RU    & Mul.Lin. & 0 \\
    INLI  & Interlife General Insurance & GR    & Mul.Lin. & 0 \\
    JDOS  & Adriatic Osiguranje & HR    & Mul.Lin. & 0 \\
    JOSVP & Groupe Josi & BE    & Reins. & 0 \\
    JUST  & Just Group & GB    & Lif.Hea. & 1 \\
    KDVORA & Dunav Osiguranje & BY    & Mul.Lin. & 0 \\
    KJUBI & Makedonija Osiguruvane & MC    & Ins.Bro. & 0 \\
    KKOSRA & Triglav Osiguranje & BY    & Mul.Lin. & 0 \\
    KMKS  & K Mk Broker Skopje & MC    & Mul.Lin. & 0 \\
    KROSRA & Krajina Osiguranje & BY    & Mul.Lin. & 0 \\
    LDA   & Linea Directa Aseguradora & SP    & Ins.Bro. & 0 \\
    LGEN  & Legal \& General & GB    & Lif.Hea. & 1 \\
    LRE   & Lancashire & GB    & Pro.Cas. & 1 \\
    LSI   & Lifestar & MB    & Lif.Hea. & 0 \\
    MAOS  & Magnat Osiguranje & YV    & Ins.Bro. & 0 \\
    MAP   & Mapfre & SP    & Mul.Lin. & 1 \\
    MINE  & Minerva Insurance & CC    & Lif.Hea. & 1 \\
    MKOSRA & Mikrofin Osiguranje & BY    & Ins.Bro. & 0 \\
    MMS   & Mapfre Middlesea & MB    & Mul.Lin. & 0 \\
    MNG   & M\&G  & GB    & Lif.Hea. & 0 \\
    MNOS  & Sava Osiguranje & YV    & Pro.Cas. & 0 \\
    MSOS  & Evroins Osiguruvane Skopje & MC    & Lif.Hea. & 0 \\
    MUV2  & Muench. Rueckvers.-Gesell. & GE    & Reins. & 1 \\
    NBG6  & Nuernberger Beteiligungs & GE    & Mul.Lin. & 1 \\
    NET   & Net Insurance & IT    & Pro.Cas. & 0 \\
    NN    & NN Group & NE    & Lif.Hea. & 1 \\
    ONDO  & Ondo Insur Tech & GB    & Pro.Cas. & 0 \\
    OPTIM & Optimco & BE    & Mul.Lin. & 0 \\
    OSPO  & Osiguritelna Polisa Skopje & MC    & Others & 0 \\
    OTIS  & Cia Internationala De Asig. As. & MK    & Mul.Lin. & 0 \\
    PANNONIA & CIG Pannonia Life & HU    & Lif.Hea. & 1 \\
    PGH   & Personal & GB    & Pro.Cas. & 1 \\
    PHNX  & Phoenix & GB    & Lif.Hea. & 1 \\
    POSR  & Pozavarovalnica Sava & SV    & Reins. & 1 \\
    PPDT  & Prva  & SV    & Others & 0 \\
    PROT  & Protector Forsikring & NO    & Pro.Cas. & 1 \\
    PRU   & Prudential & HK    & Lif.Hea. & 1 \\
    PYDR  & Paydrive & SW    & Ins.Bro. & 0 \\
    PZU   & Powszechny Zaklad Ubezpiec. & PD    & Pro.Cas. & 1 \\
    RAYSG & Ray Sigorta & TU    & Lif.Hea. & 1 \\
    RENI  & Renaissance Insurance Group & RU    & Mul.Lin. & 0 \\
    RLV   & Rheinland Holding & GE    & Mul.Lin. & 1 \\
    SAMPO & Sampo Oyj & FI    & Mul.Lin. & 1 \\
    SBRE  & Sabre & GB    & Pro.Cas. & 0 \\
    SCR   & SCOR  & FR    & Reins. & 1 \\
    SFAB  & Solid Forsakring & SW    & Pro.Cas. & 0 \\
    SGS   & Cash Life & GE    & Ins.Bro. & 1 \\
    SJOVA & Sjova-Almennar Tryggingar & IC    & Mul.Lin. & 1 \\
    SKUN  & Universalna & UK    & Mul.Lin. & 0 \\
    SLHN  & Swiss Life Holding & SZ    & Lif.Hea. & 1 \\
    SLP401E & Allianz-Slovenska Poistovna & SO    & Mul.Lin. & 0 \\
    SODJRA & Wiener Osiguranje Vienna & BY    & Pro.Cas. & 0 \\
    SORN  & Oranta & UK    & Mul.Lin. & 0 \\
    SOSOR & Sarajevo Osiguranje & BY    & Mul.Lin. & 0 \\
    SREN  & Swiss Re & SZ    & Reins. & 1 \\
    STB   & Storebrand & NO    & Mul.Lin. & 1 \\
    SWIO  & Grawe Nezivotno Osiguranje & YV    & Pro.Cas. & 0 \\
    TBK   & Transilvania Broker de Asig. & RO    & Ins.Bro. & 0 \\
    TBKO  & Sava Osiguruvanje & MC    & Pro.Cas. & 0 \\
    TLX   & Talanx & GE    & Reins. & 1 \\
    TM    & Tm Hf & IC    & Pro.Cas. & 0 \\
    TOP   & Topdanmark & DE    & Pro.Cas. & 1 \\
    TRYG  & Tryg  & DE    & Pro.Cas. & 1 \\
    TURSG & Turkiye Sigorta & TU    & Pro.Cas. & 1 \\
    UNI   & Unipol Gruppo & IT    & Mul.Lin. & 1 \\
    UQA   & UNIQA & AS    & Mul.Lin. & 1 \\
    US    & Unipolsai Assicurazioni & IT    & Mul.Lin. & 1 \\
    VAHN  & Vaudoise Assurances Holding & SZ    & Pro.Cas. & 1 \\
    VIG   & Vienna Insurance Group & AS    & Mul.Lin. & 1 \\
    VROS  & Vardar Osiguruvane & MC    & Pro.Cas. & 0 \\
    WTW   & Willis Towers Watson & GB    & Ins.Bro. & 1 \\
    WUW   & Wuestenrot \& Wuerttemberg. & GE    & Lif.Hea. & 1 \\
    ZEOS  & Uniqa Zivotno Osiguranje & YV    & Lif.Hea. & 0 \\
    ZURN  & Zurich & SZ    & Mul.Lin. & 1 \\
    ZVTG  & Zavarovalnica Triglav & SV    & Lif.Hea. & 1 \\
	%
	\bottomrule
\multicolumn{5}{@{}m{\textwidth}@{}}{
\textit{Notes}: From left to right, the table reports the following information on the insurance companies included in our sample: (i) the company's abbreviated name (Label); (ii) the full company name (Company); (iii) the country of origin (Country); 
(iv) the insurance subsector to which the company belongs (Subsector), considering insurance brokers (Ins.Bro.), life health (Lif.Hea.), multiline (Mul.Lin.), property and casualty (Pro.Cas.), and reinsurance (Reins.); and (v) a dummy variable (Network) which takes the value of one if the company meets the data availability and liquidity requirements to be included in the network analysis described in Figures \ref{fig:CompaniesNetwork_LogRet}\textemdash \ref{fig:CompaniesNetwork_CARES}, and the value of zero otherwise. 
}\\
\end{longtable}
\end{footnotesize}

\begin{table}[htbp]
\centering
\begin{tabular}{lll}
\toprule
\multicolumn{3}{c}{Percentile threshold} \\
\midrule
60th & 75th & 90th \\
\midrule
Aegon & Aegon & \textbf{Aviva} \\
Ageas & Ageas & \textbf{Legal \& General} \\
Allianz & \textbf{Aviva} & \textbf{Prudential} \\
\textbf{Aviva} & AXA & \\
AXA & Generali & \\
Baloise Holding & \textbf{Legal \& General} & \\
Chubb & \textbf{Prudential} & \\
Generali & Storebrand & \\
Hannover Rueck & & \\
Helvetia Holding & & \\
\textbf{Legal \& General} & & \\
Muench. Rueckvers.-Gesell. & & \\
\textbf{Prudential} & & \\
Storebrand & & \\
Swiss Life Holding & & \\
Swiss Re & & \\
Vaudoise Assurances Holding & & \\
Willis Towers Watson & & \\
Zurich & & \\
\bottomrule
\end{tabular}
\caption{Systematically central insurers under different thresholds. 
This table reports the insurance companies identified as central when applying three alternative cutoff values (60th, 75th, and 90th percentiles) to the distribution of link absolute values. 
Companies that are consistently identified across all threshold values are shown in bold.}\label{tab:robust_different_threshold}
\end{table}

\begin{table}[htbp]
  \centering
  \begin{footnotesize}
    \begin{tabular}{L{2cm}C{2.4cm}C{2.4cm}C{2.4cm}C{2.4cm}}    
    \hline
 & Log-Return & Log-Volatility & CAViaR & CARES \\
\toprule
\multicolumn{5}{c}{\textbf{Years 2013--2015}} \\
\midrule
Insurance & 67.29 & 61.25 & 65.84 & 64.08 \\
Equity    & 57.41 & 41.27 & 58.47 & 51.91 \\
Bank      & 59.61 & 49.09 & 48.72 & 44.66 \\
Bond      &  8.01 &  6.24 &  6.14 & 11.76 \\
\midrule
\multicolumn{5}{c}{\textbf{Years 2016--2018}} \\
\midrule
Insurance & 67.92 & 54.72 & 58.16 & 64.00 \\
Equity    & 46.16 & 36.48 & 55.23 & 43.94 \\
Bank      & 66.78 & 59.27 & 66.12 & 64.81 \\
Bond      & 15.13 & 19.53 & 13.11 & 15.42 \\
\midrule
\multicolumn{5}{c}{\textbf{Percentage Variation (\%)}} \\
\midrule
Insurance &  0.93 &  -10.66 &  -11.66 &  -0.13 \\
Equity    &  -19.61 &  -11.60 &  -5.54 &  -15.36 \\
Bank      &  12.04 &  20.74 &  35.71 &  45.10 \\
Bond      &  88.83 &  213.19 &  113.33 &  31.04 \\
\bottomrule
\end{tabular}
 \caption{Systemic risk contributions before and after Solvency II (multi-year estimates). This table reports the contribution to others of the insurance, equity, bank and bond markets estimated in years 2013 to 2015 (Panel A) and in years 2016 to 2018 (Panel B). The contribution to others is reported for the log-return, the log-volatility, the CAViaR and the CARES risk measures. The insurance market regulation Solvency II was introduced January 1, 2016. Panel C reports the percentage variation in the contribution to others between the 2013--2015 and 2016--2018 periods.}
\label{tab:contribution_to_others_multi}
  \end{footnotesize}
\end{table}%

\begin{figure}[hbt!]
\hspace{-0.4cm}
\includegraphics[scale=0.32]{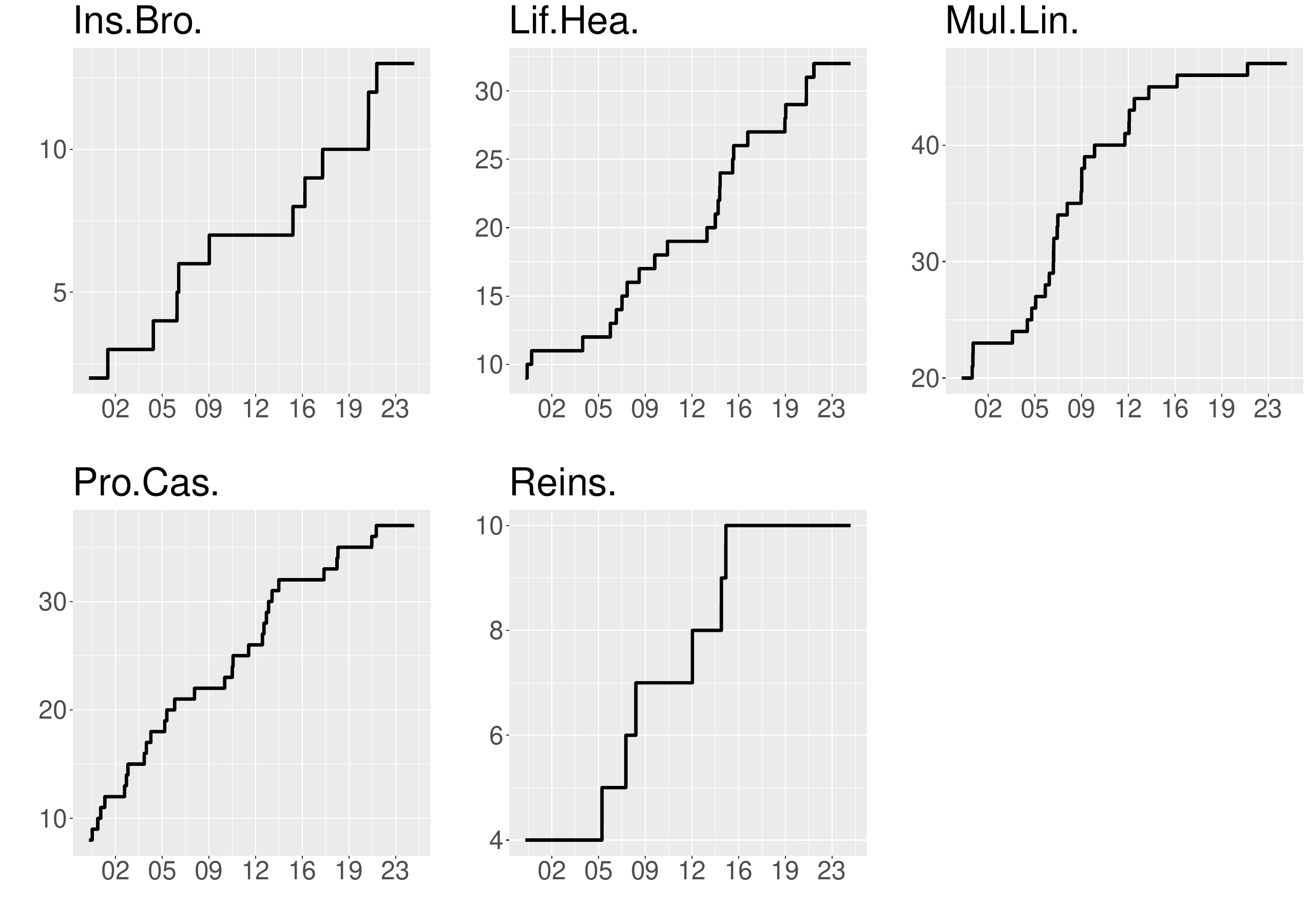}
\caption{This figure plots the evolution of the number of individual European insurance companies in our sample by subsector. The subsectors are: insurance brokers (Ins.Bro.), life health (Lif.Hea.), multiline (Mul.Lin.), property and casualty (Pro.Cas.), and reinsurance (Reins.). The sample goes from January 3, 2000 to October 22, 2024. Data are from Bloomberg.}
    \label{fig:activepositions}
\end{figure}

\begin{figure}[hbt!]
\hspace{-0.4cm}
\includegraphics[scale=0.32]{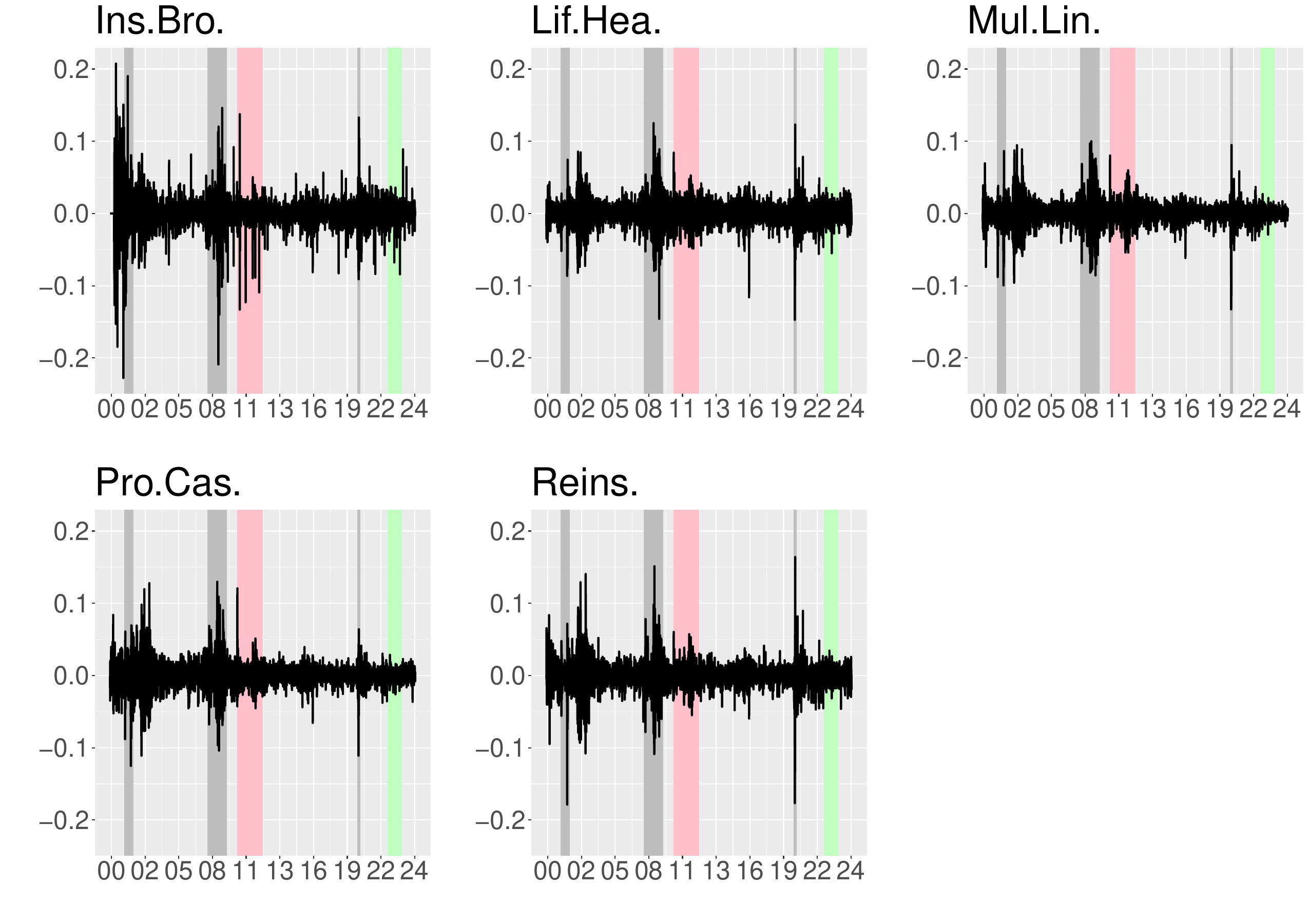}
\caption{This figure plots the logged daily returns the following subsectors of the insurance market: insurance brokers (Ins.Bro.), life health (Lif.Hea.), multiline (Mul.Lin.), property and casualty (Pro.Cas.), and reinsurance (Reins.). The sample goes from January 3, 2000 to October 22, 2024. Data are from Bloomberg.}
    \label{fig:portfolios_logreturns}
\end{figure}

\begin{figure}[hbt!]
\hspace{-0.4cm}
\includegraphics[scale=0.32]{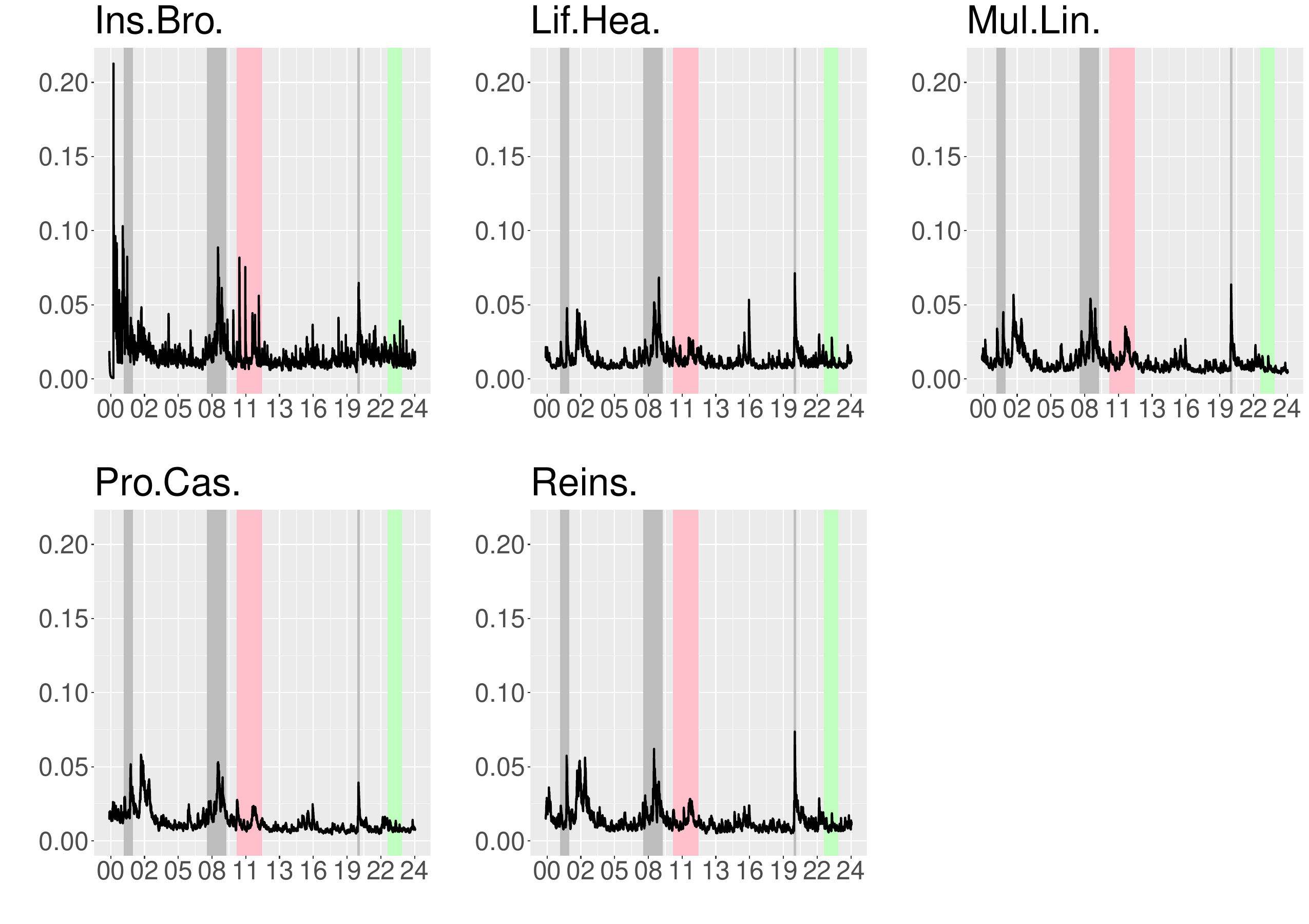}
\caption{This figure plots the daily conditional volatility of the following subsectors of the insurance market: insurance brokers (Ins.Bro.), life health (Lif.Hea.), multiline (Mul.Lin.), property and casualty (Pro.Cas.), and reinsurance (Reins.). The sample goes from January 3, 2000 to October 22, 2024.}
    \label{fig:portfolios_logcvol}
\end{figure}

\begin{figure}[hbt!]
\hspace{-0.4cm}
\includegraphics[scale=0.32]{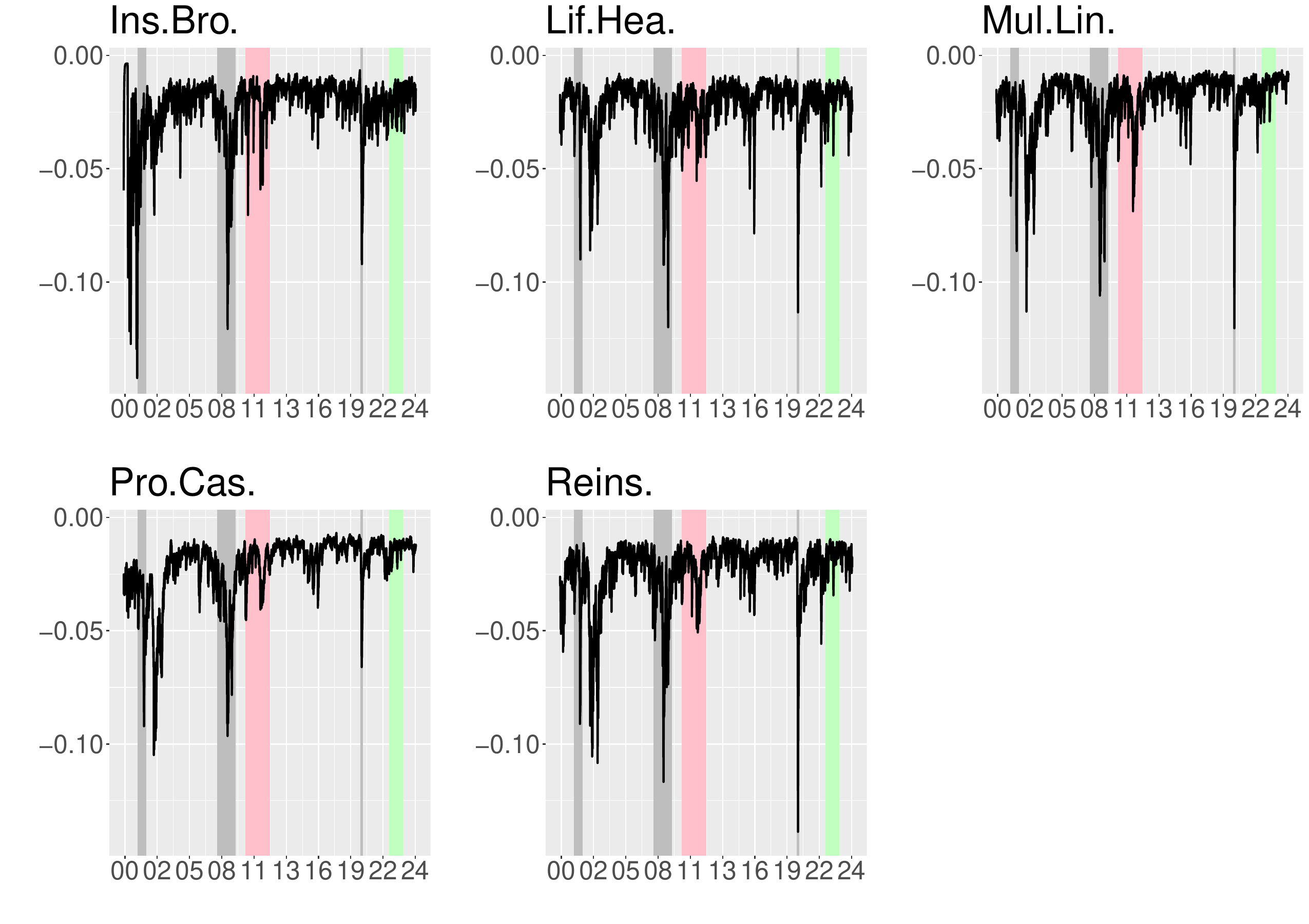}
\caption{This figure plots the daily conditional value-at-risk estimated by the CAViaR model of the following subsectors of the insurance market: insurance brokers (Ins.Bro.), life health (Lif.Hea.), multiline (Mul.Lin.), property and casualty (Pro.Cas.), and reinsurance (Reins.). The sample goes from January 3, 2000 to October 22, 2024.}
    \label{fig:portfolios_CAViaR}
\end{figure}

\begin{figure}[hbt!]
\hspace{-0.4cm}
\includegraphics[scale=0.32]{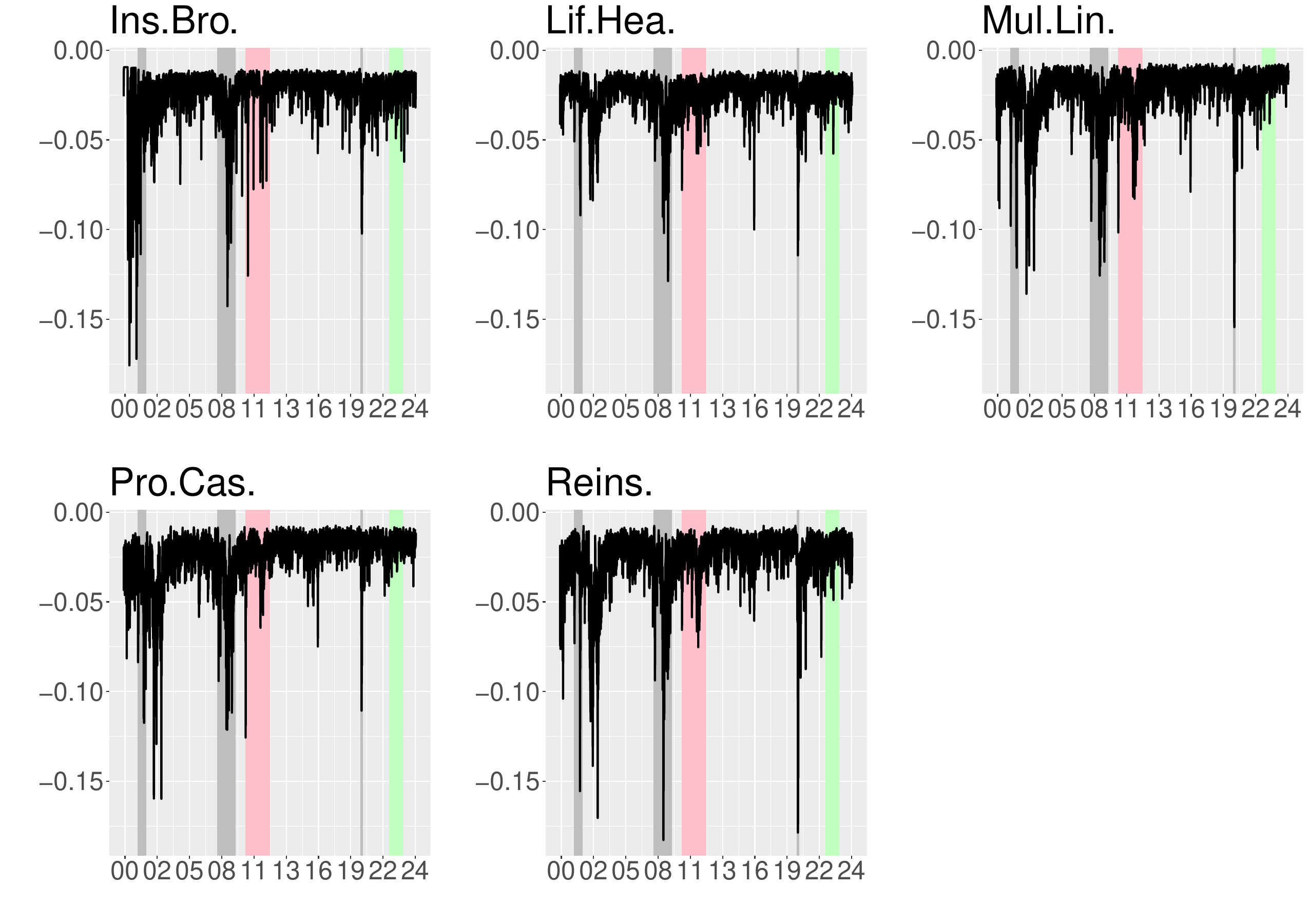}
\caption{This figure plots the daily conditional expected shortfall estimated by the CARES model of the following subsectors of the insurance market: insurance brokers (Ins.Bro.), life health (Lif.Hea.), multiline (Mul.Lin.), property and casualty (Pro.Cas.), and reinsurance (Reins.). The sample goes from January 3, 2000 to October 22, 2024.}
    \label{fig:portfolios_cares}
\end{figure}

\begin{table}[!htbp]
\centering
\small
\resizebox{\textwidth}{!}{%
\begin{tabular}{lcccccc}
\hline
 & \multicolumn{3}{c}{Pooled} & \multicolumn{3}{c}{TWFE} \\
 & Bank & Bond & Equity & Bank & Bond & Equity \\
\hline
\multicolumn{7}{l}{\textbf{Panel A: CARES}} \\
Lev & 0.342*** & 0.081** & 0.307*** & 0.015 & -0.039* & -0.031 \\
 & (0.124) & (0.033) & (0.100) & (0.075) & (0.021) & (0.073) \\
Home Bias & 22.075*** & 5.381*** & 16.889*** & -10.667 & -9.880 & -16.157 \\
 & (5.129) & (1.207) & (4.172) & (35.345) & (7.369) & (30.062) \\
Sov Spread & 2.059** & 0.812*** & 3.577*** & -2.543*** & 0.136 & -0.566 \\
 & (0.841) & (0.250) & (0.970) & (0.905) & (0.194) & (1.036) \\
Home Bias $\times$ Sov Spread & -5.629*** & -1.877*** & -8.136*** & 3.936** & -0.240 & 0.926 \\
 & (1.972) & (0.568) & (2.246) & (1.718) & (0.412) & (1.851) \\
N & 33882 & 33882 & 33882 & 33882 & 33882 & 33882 \\
Firm FE & No & No & No & Yes & Yes & Yes \\
Time FE & No & No & No & Yes & Yes & Yes \\
\hline
\multicolumn{7}{l}{\textbf{Panel B: Log-Ret}} \\
Lev & 0.372** & 0.097** & 0.313*** & -0.072 & -0.025 & -0.063 \\
 & (0.148) & (0.048) & (0.117) & (0.074) & (0.030) & (0.048) \\
Home Bias & 24.331*** & 6.656*** & 23.822*** & 14.784 & 8.231 & 31.109*** \\
 & (8.563) & (2.091) & (7.000) & (27.507) & (5.852) & (10.307) \\
Sov Spread & 2.427*** & 0.851*** & 4.024*** & -1.154 & -0.268 & -0.429 \\
 & (0.887) & (0.282) & (1.009) & (0.760) & (0.337) & (0.672) \\
Home Bias $\times$ Sov Spread & -6.544*** & -1.786*** & -9.963*** & 2.321 & 0.560 & 0.697 \\
 & (2.203) & (0.669) & (2.412) & (1.451) & (0.681) & (1.237) \\
N & 33236 & 33236 & 33236 & 33236 & 33236 & 33236 \\
Firm FE & No & No & No & Yes & Yes & Yes \\
Time FE & No & No & No & Yes & Yes & Yes \\
\hline
\end{tabular}
}%
\caption{Dependent variable: directional spillover from insurance to Bank, Bond, and Equity markets. Robust standard errors are one-way clustered at country level.}
\label{tab:twopanel_logret_cares_SE_country_clustered}
\end{table}

\end{document}